%% file: main_v3.tex
\documentclass[12pt]{article}

\usepackage{amsmath, amssymb, amsthm, amsfonts}
\usepackage{graphicx}
\usepackage{tikz}
\usetikzlibrary{shapes.geometric, arrows}
\usepackage[T1]{fontenc}
\usepackage[utf8]{inputenc}
\usepackage[english]{babel}
\usepackage{csquotes}
\usepackage[round]{natbib}
\usepackage{hyperref}
\usepackage{url}
\usepackage{booktabs}
\usepackage{tabularx}
\usepackage{multirow}
\usepackage[expansion=false]{microtype}
\usepackage{caption}
\usepackage{float}
\usepackage{xcolor}
\usepackage{algorithm}
\usepackage{algpseudocode}
\algnewcommand\algorithmicinput{\textbf{Input:}}
\algnewcommand\algorithmicoutput{\textbf{Output:}}
\algnewcommand\Input{\item[\algorithmicinput]}
\algnewcommand\Output{\item[\algorithmicoutput]}

\title{Toward Epistemic Stability: Engineering Consistent Procedures for Industrial LLM Hallucination Reduction}
\author{Brian Freeman$^{*}$ \and Adam Kicklighter \and Matt Erdman \and Zach Gordon\\[0.5ex]
        \small Trane Technologies, Davidson NC, USA\\[0.25ex]
        \small $^{*}$Corresponding author: \texttt{brian.freeman@tranetechnologies.com}}
\date{March 2026}

\begin{document}
\maketitle


\begin{abstract}
Hallucinations in large language models (LLMs) are outputs that are syntactically coherent but factually incorrect or contextually inconsistent. They are persistent obstacles in high-stakes industrial settings such as engineering design, enterprise resource planning, and IoT telemetry platforms. We present and compare five prompt engineering strategies intended to reduce the variance of model outputs and move toward repeatable, grounded results without modifying model weights or creating complex validation models.  These methods include: (M1)~\emph{Iterative Similarity Convergence}, (M2)~\emph{Decomposed Model-Agnostic Prompting}, (M3)~\emph{Single-Task Agent Specialization}, (M4)~\emph{Enhanced Data Registry}, and (M5)~\emph{Domain Glossary Injection}. Each method is evaluated against an internal baseline using an LLM-as-Judge framework over 100 repeated runs per method (same fixed task prompt, stochastic decoding at $\tau = 0.7$. Under this evaluation setup, M4 (Enhanced Data Registry) received ``Better'' verdicts in all 100 trials; M3 and M5 reached 80\% and 77\% respectively; M1 reached 75\%; and M2 was net negative at 34\% when compared to single shot prompting with a modern foundation model. We then developed enhanced version 2 (v2) implementations and assessed them on a 10-trial verification batch; M2 recovered from 34\% to 80\%, the largest gain among the four revised methods. We discuss how these strategies help overcome the non-deterministic nature of LLM results for industrial procedures, even when absolute correctness cannot be guaranteed. We provide pseudocode, verbatim prompts, and batch logs to support independent assessment.
\end{abstract}

\section*{List of Acronyms and Variables}

\subsection*{Acronyms}
\begin{tabular}{@{}ll@{}}
AHU  & Air Handling Unit \\
AI   & Artificial Intelligence \\
API  & Application Programming Interface \\
BAS  & Building Automation System \\
BMS  & Building Management System \\
CFM  & Cubic Feet per Minute \\
DDC  & Direct Digital Controller \\
DX   & Direct Expansion (refrigerant-based cooling) \\
ERP  & Enterprise Resource Planning \\
EWT  & Entering Water Temperature \\
FCU  & Fan Coil Unit \\
HVAC & Heating, Ventilation, and Air Conditioning \\
IoT  & Internet of Things \\
LLM  & Large Language Model \\
LWT  & Leaving Water Temperature \\
MAT  & Mixed Air Temperature \\
OAT  & Outside Air Temperature \\
RAG  & Retrieval-Augmented Generation \\
RLHF & Reinforcement Learning from Human Feedback \\
RTU  & Rooftop Unit \\
SAT  & Supply Air Temperature \\
VAV  & Variable Air Volume \\
VFD  & Variable Frequency Drive \\
\end{tabular}

\subsection*{Variables and Mathematical Notation}
\begin{tabular}{@{}ll@{}}
$p$       & Per-item base accuracy (probability a single prediction is correct) \\
$n$       & Number of elements in an output sequence \\
$k$       & Number of elements verified as epistemically certain \\
$\tau$    & Temperature parameter for LLM token sampling \\
$\sigma_{\text{sim}}$ & Semantic similarity threshold used in M1 convergence check \\
$z_i$     & Logit score for token $i$ before softmax \\
$p_i$     & Probability of selecting token $i$ after temperature scaling \\
$\hat{A}_{\text{train}}$ & Empirical accuracy on training set \\
$\hat{A}_{\text{test}}$  & Empirical accuracy on test/validation set \\
$\mathcal{L}_{\text{CE}}$ & Cross-entropy loss function \\
$N_{\text{train}}$ & Number of samples in training set \\
$N_{\text{test}}$  & Number of samples in test set \\
$y_i$     & True label or target output for sample $i$ \\
$\hat{y}_i$ & Model's predicted output for sample $i$ \\
$\mathbb{I}[\cdot]$ & Indicator function (1 if condition is true, 0 otherwise) \\
$\theta$  & Model parameters \\
$w_t$     & Token at position $t$ \\
$L$       & Sequence length (number of token positions) \\
\end{tabular}

\section{Introduction}
\label{sec:intro}

Large Language Models (LLMs) and agentic AI systems are increasingly deployed in industrial settings to help produce maintenance procedures, inspection checklists, engineering summaries, telemetry pipeline plans, and incident response reports ~\citep{raza2025}. Deployment, however, has not consistently translated into sustained adoption or met early user expectations \citep{dilmegani2025}. One well-documented barrier is reliability: many industrial tasks require outputs that are not merely plausible on average, but correct for a specific instance and auditable after the fact \citep{shah2024}.

A central challenge is structural. LLM outputs are generated probabilistically and are not deterministically grounded in an authoritative source of truth at inference time. As a result, LLMs can produce hallucinations-content that appears credible but is factually incorrect and is particularly hazardous when output is acted upon without independent review
\citep{ji2023}. This problem compounds for multi-element outputs even if each element is individually likely to be correct, the probability that an entire $n$-element artifact is fully correct can be much lower than practitioners expect.

Existing mitigation strategies broadly divide into three families: (i)~training-time interventions such as reinforcement learning from human feedback (RLHF)~\citep{ouyang2022} and direct preference optimization (DPO)~\citep{rafailov2023}, (ii)~inference-time augmentation such as retrieval-augmented generation (RAG)~\citep{lewis2021}, and (iii)~prompt engineering, which reshapes or decomposes the input given to a fixed model~\citep{white2023}. Training-time approaches require access to model internals and significant compute. RAG depends on a high-quality retrieval corpus. Prompt engineering remains the most practical option for practitioners working with hosted, black-box API models, and we focus exclusively on this family.

\subsection{Temperature and Sampling Variance}
\label{sec:temperature}

Temperature controls sampling randomness during LLM decoding. Technically, temperature $\tau$ scales logits $z_i$ prior to softmax \citep{xuan2025}:

\begin{equation}
\label{eq:temperature_softmax}
p_i = \frac{\exp\!\left(\dfrac{z_i}{\tau}\right)}%
           {\sum_{j} \exp\!\left(\dfrac{z_j}{\tau}\right)}.
\end{equation}

Lower $\tau$ sharpens the distribution and makes outputs more repeatable; higher $\tau$ increases diversity \citep{Li2025,castillo2025}. In industrial settings, lower temperature can reduce run-to-run variability, but it cannot prevent hallucinations caused by missing, outdated, or incorrect associations encoded in model parameters. We therefore do not treat temperature adjustment as a hallucination-reduction strategy in this work, and all task runs use the API default unless otherwise noted.

\subsection{Hallucinations in Large Language Models}

An hallucination in an LLM refers to generated content that is syntactically coherent and contextually plausible but factually inaccurate or fabricated ~\citep{alansari2025}. They are operationally dangerous because they tend to match domain conventions, appearing credible to reviewers who lack independent ground truth.

Several factors contribute to hallucinations in industrial deployments. Domain coverage is often limited: proprietary or specialist data may be underrepresented in training corpora, so the model fills gaps by analogy with related but distinct contexts \citep{singh2025, kalai2025}. Similarly, in the absence of explicit grounding, models can infer component tolerances, fault codes, or inter-system relationships from statistical co-occurrence patterns rather than authoritative sources \citep{Kim2025,reizinger2024}. These failure modes are not reliably visible from model benchmarks, which measure average accuracy on held-out sets rather than per-instance correctness.

\subsection{Error in Statistical Machine Learning Systems}

Supervised models are trained to minimize average empirical loss across a dataset, not to be correct on any specific inference request \citep{jiang2020,vapnik1991}. For LLMs, training minimizes cross-entropy over token sequences:

\begin{equation}
\label{eq:cross_entropy_loss}
\mathcal{L}_{\text{CE}}
  = - \sum_{t=1}^{L} \log p_{\theta}(w_t \mid w_{<t}),
\end{equation}

rewarding statistically probable continuations rather than per-instance factual accuracy \citep{he2025,elharrouss2025}. The gap between aggregate training behavior and deployment correctness on a specific industrial query is the direct motivation for the inference-time strategies in this paper.

\subsection{Handling Errors}
\label{sec:handling_errors}

In deployment, ground truth is typically unavailable for each new prediction instance, so reported test-set performance provides only an expected error rate rather than a per-instance correctness guarantee.

For discrete, multi-element outputs-for example, a bill of materials or a set of acceptance criteria is used to treat each element as independently correct with probability $p$. The probability that an entire $n$-item sequence is fully error-free is:

\begin{equation}
\label{eq:prob_all_correct}
P(\text{all correct}) = p^{n}.
\end{equation}

With $p = 0.9$ and $n = 10$, this gives $P \approx 0.35$: even at 90\% per-item accuracy, there is roughly a two-in-three chance that the full output contains at least one error. The expected number of errors within a sequence of operations is:

\begin{equation}
\label{eq:expected_errors}
E[\text{errors}] = n(1-p).
\end{equation}

Without verification we cannot identify which elements are wrong. This is the practical argument for inference-time strategies that reduce individual error probability before output is assembled. The formula $P(\text{all correct}) = p^n$ assumes each element fails independently of every other. 
LLM outputs do not satisfy this assumption: when a model misinterprets a constraint early in a multi-step response, all downstream elements that depend on that constraint tend to fail together \citep{ifty2026}. Under this correlated failure pattern the true probability of a fully correct output can be lower than $p^n$ when errors cluster (one wrong assumption produces a cascade), or higher when a single correct reasoning step constrains many later elements. The formula is therefore better understood as an intuition pump for the compounding risk of multi-element outputs than as a precise probability bound. We do not estimate the empirical correlation structure of LLM errors in this work.

\subsection{Application-Specific Failure Modes}

Failure modes that directly motivated this project were all observed in a production Azure OpenAI deployment querying HVAC and BMS sensor data and supporting enterprise software incident response:

\begin{enumerate}
  \item \emph{Acronym ambiguity}: domain abbreviations such as AHU, MAT, VFD,
        and EWT were interpreted inconsistently across separate query sessions, reflecting the model's non-deterministic mapping of polysemous terms when no explicit glossary is provided. This often produced diagnostics that referenced the wrong physical system component.
  \item \emph{Context collapse}: complex multi-requirement prompts (e.g.,
        multi-sprint project plans with Fibonacci estimates, role assignments, and inter-sprint dependencies) caused the model to silently drop constraints, yielding structurally complete but incomplete deliverables.
  \item \emph{Cascading inconsistency}: when a single agent handled multiple analytical sub-tasks in one call, an error in an early sub-task propagated into later outputs, producing a coherent but wrong post-mortem.
\end{enumerate}

Each failure mode maps directly to one or more of the five methods evaluated in this paper.

\subsection{Contributions and Scope}

This paper makes the following contributions:

\begin{itemize}
  \item A systematic comparison of five prompt engineering strategies using a consistent internal-baseline, LLM-as-Judge framework over 100 repeated runs per method (same prompt, stochastic decoding).
  \item Enhanced v2 implementations of four methods addressing specific
        weaknesses identified in the baseline evaluation, assessed on a 10-trial verification batch.
  \item Verbatim prompts (Appendix), pseudocode, and batch results for all
        methods, enabling replication and independent assessment.
  \item An honest analysis of where this evaluation design has limits: same
        model as generator and judge, narrow task set, coarse verdict scale.
\end{itemize}

The study is explicitly scoped to one model family (Azure OpenAI GPT-5), four distinct task scenarios (Table~\ref{tab:scenarios}), and an LLM-as-Judge rubric driven by the same model. Results should be taken as directional signals in our deployment context, not as general benchmarks.

The paper is organized as follows. Section~\ref{sec:epistemic} frames the concept of epistemic certainty as a design lens. Section~\ref{sec:related} reviews related work. Section~\ref{sec:methodology} describes the five baseline methods. Section~\ref{sec:setup} details experimental setup and evaluation validity. Section~\ref{sec:enhanced} presents the v2 implementations. Section~\ref{sec:results} reports results. Section~\ref{sec:discussion} discusses findings and limitations. Section~\ref{sec:conclusion} concludes.

\section{Epistemic Stability: Moving Toward Operable Certainty}
\label{sec:epistemic}

Epistemic certainty-in its strongest philosophical form-refers to a state of absolute justified conviction regarding a proposition \citep{Stanley2008}. In AI terms, Alvarado argues that AI methods are fundamentally epistemic technologies: tools designed to support inquiry, analysis, and the manipulation of knowledge representations \citep{alvarado2023}.  Coeckelbergh frames epistemic agency in LLMs around how models influence human belief formation \citet{Coeckelbergh2025}. Spivack treat epistemic uncertainty as a signal of missing training coverage \citep{spivack2025}.

In industrial applications where non-deterministic models represent operation risk, we introduce the concept of \textbf{Epistemic Stability}. While certainty remains an aspirational ideal, stability is the degree to which an engineering procedure yields consistent, repeatable, and defensible results across multiple runs. For a human engineer, a system that produces a correct diagnosis 95\% of the time but provides a radically different (and wrong) explanation on the next identical request is not "stable" enough for deployment in high-stakes environments.

We do not claim that LLM outputs can be made epistemically certain in the philosophical sense. Instead, we use the term operationally to mean \textit{the degree to which the inputs, data enrichments, and domain grounding signals provided at inference time are independently checkable and well-justified}. A response is closer to being epistemically grounded when the context supplied to the model is accurate, structured, and independently verifiable-even when the model's own reasoning cannot be.

This framing motivates a simple test: for a given output, could a domain expert trace its key claims back to the provided context? If yes, the response is more likely to be reliable. If the model had to reach beyond the supplied context to fill gaps, hallucination risk increases. We formalize this as:

\begin{equation}
\label{eq:binary_certainty}
C(q) =
\begin{cases}
1, & \text{if the key claims in the output for query $q$ are traceable} \\
   & \text{to the provided context or independently verifiable facts} \\
0, & \text{otherwise.}
\end{cases}
\end{equation}

In industrial environments, the pursuit of $C(q) = 1$ is driven by safety and economic stakes. For instance, an incorrect diagnosis in a BMS that leads to the short-cycling of a centrifugal chiller carries high risk of premature bearing failure and significant maintenance costs and damage to the provider's reputation. 

\subsection{Machine Learning Context}

The gap between average training behavior and per-instance deployment correctness means we cannot rely on aggregate benchmark accuracy as a guarantee for any individual HVAC diagnostic or ERP incident report \citep{Gromov2023,Aliferis2024}. For a given query $q$, let $P_{\text{LLM}}(q)$ denote the model's per-instance correctness probability and let $V(q)$ denote the event that the output is independently verified before use. A downstream policy that replaces unverified incorrect outputs would achieve:

\begin{equation}
\label{eq:verification_correctness}
P_{\text{correct}}(q)
  = P\!\left(V(q)\right)
    + \left(1 - P\!\left(V(q)\right)\right)\cdot P_{\text{LLM}}(q).
\end{equation}

The five strategies in this paper each attempt to increase $P_{\text{LLM}}(q)$ directly, without requiring $P(V(q)) > 0$: M1 by exploiting output stability as a proxy; M2 and M3 by narrowing the scope of each individual LLM call; and M4 and M5 by improving the quality of the input; .

\subsection{Scope of the Epistemic Certainty Claim}

We note explicitly that the label ``epistemic certainty'' is aspirational in this context. No LLM output-regardless of prompt design-can be epistemically certain in the philosophical sense. We use it as a design goal: build the context so well that an external reviewer could check every claim without needing the model to explain itself. When prompts supply that kind of grounding and scope, we observe better verdicts from our judge. We do not claim that better judge verdicts necessarily mean ground-truth correctness.

\section{Related Work}
\label{sec:related}

\subsection{Hallucination Taxonomy and Surveys}

\citet{ji2023} provides a comprehensive taxonomy distinguishing
intrinsic (contradicting source material) from extrinsic
(unverifiable fabrication) hallucinations. \citet{huang2025} systematically categorize mitigation strategies across data, training, and inference stages. Our work operates exclusively at inference time.

\subsection{Self-Consistency and Iterative Prompting}

\citet{wang2023} show that sampling multiple reasoning paths
and aggregating by majority vote improves accuracy on reasoning benchmarks. M1~v1 extends this to arbitrary text generation by using semantic similarity across independently generated responses as the convergence signal rather than a majority-vote count.

\subsection{Chain-of-Thought and Decomposition}

\citet{wei2023} demonstrate that step-by-step reasoning using chain-of-thought prompting
substantially reduces factual errors on multi-step tasks. Least-to-most prompting decomposes queries into ordered sub-problems \citep{zhou2023}. M2 applies decomposition at the data-extraction layer rather than the reasoning layer by extracting facts from prose synthesis. We describe this as ``model-agnostic synthesis'' in the sense that the synthesis step receives structured facts rather than the raw prompt. In practice, quality still depends on what the extractor captured (see the fuller discussion in Section~\ref{subsec:m2}).

\subsection{Self-Critique and Reflection}

\citet{madaan2023} introduce Self-Refine, iterating on a draft until quality criteria are met. \citet{shinn2023} extend this with verbal reinforcement signals. M1~v2 constrains the critique to exactly three named flaws, providing a deterministic exit condition and making the critique checkable by a reader of the appendix (see Appendix~\ref{sec:appendix-m1v2}).

\subsection{Multi-Agent Architectures}

\citet{wu2023} and related frameworks orchestrate multiple LLM agents.
\citet{park2023} demonstrated emergent collaborative behavior in agent
societies. M3 tests a specific form of agent specialization: four sequential single-task agents versus one multi-task agent handling all four analytical roles simultaneously. The M3~v2 Reconciler is similar in spirit to critic-actor architectures \citep{yuan2025}, but operates on already-completed outputs rather than iterative rollouts.

\subsection{Knowledge-Augmented Prompting}

RAGs augment prompts with retrieved documents, improving factual grounding in open-domain tasks relying on vector-based similarity searches \citep{lewis2021}. For narrow industrial domains, however, maintaining a retrieval index is impractical due to redundant meanings of words and acronyms. Relevant context is a function of the specific device registry or vocabulary in use, which varies by domain. M4 and M5 follow a lighter-weight approach: structured context is assembled at request time and injected directly into the prompt. M4's registry schema and M5's controlled glossary are both domain artifacts that could not be extracted from a general-purpose index. M5~v2 adds a dynamic selection step to inject only query-relevant terms, similar to the selective augmentation studied in \citet{shi2023}.

\section{Methodology: Five Baseline Strategies}
\label{sec:methodology}

\begin{figure}[htbp]
  \centering
  \includegraphics[width=0.95\textwidth]{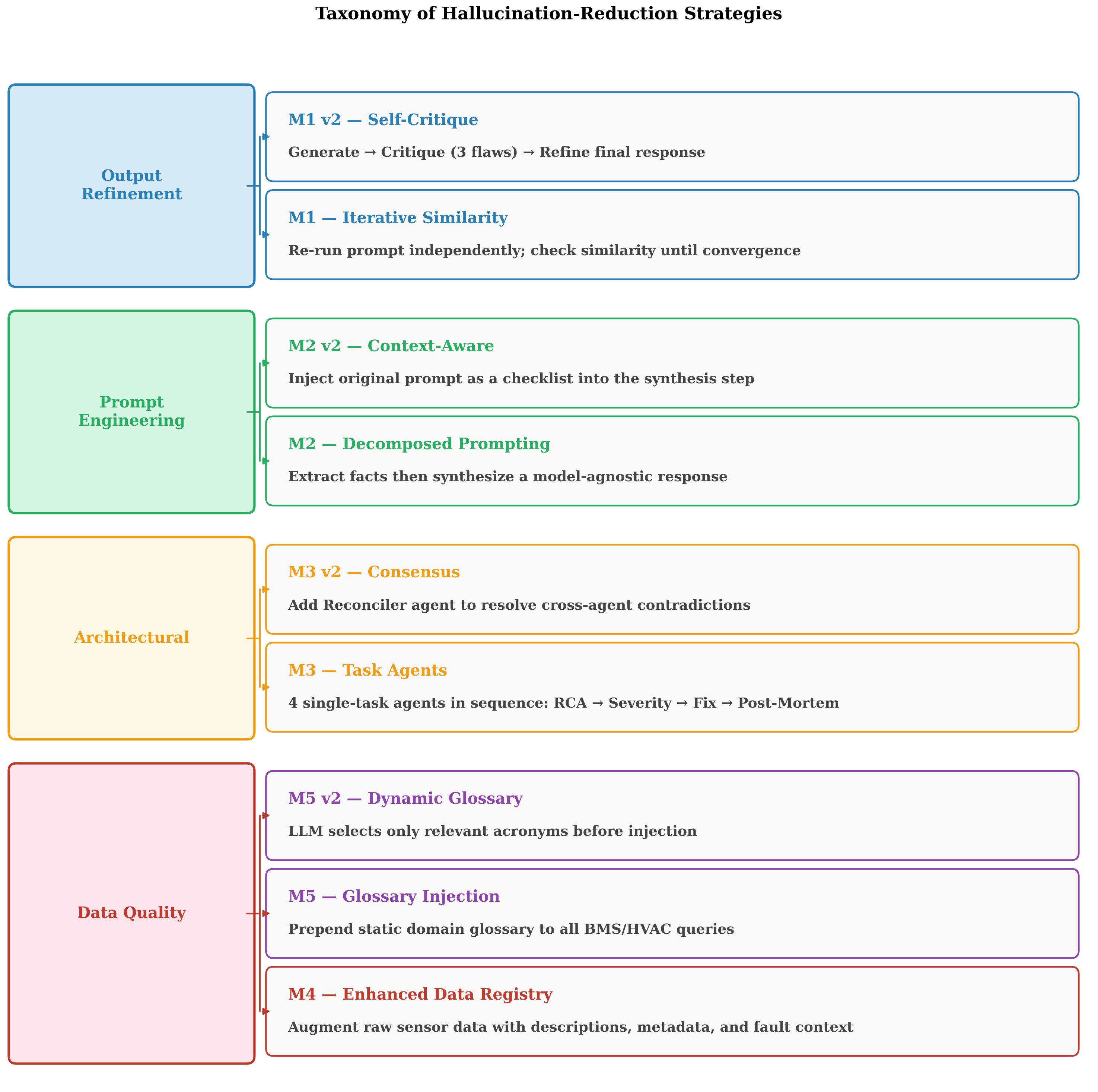}
  \caption{Taxonomy of the five hallucination-reduction strategies and the
root cause each addresses. M1 and M2 target prompt reasoning and structure; M3 targets agent architecture; M4 and M5 target input data quality.}
  \label{fig:taxonomy}
\end{figure}

All five methods share one structural choice: each method generates its own
internal baseline response using an unmodified prompt, then applies its enhancement, and the judge compares those two outputs. The primary objective of these enhancements is not to guarantee absolute factual correctness-which remains a challenge given the probabilistic nature of transformer-based models-but to provide consistent, repeatable results that overcome the non-deterministic drift typical of raw LLM outputs. Figure~\ref{fig:taxonomy} shows how the five methods group by the root cause they target: M1 (Iterative Convergence), M2 (Decomposed Prompting), M3 (Agent Specialization), M4 (Enhanced Data Registry), and M5 (Glossary Injection).

\subsection{M1: Iterative Similarity Convergence}
\label{subsec:m1}

M1 begins from a practical observation: when the same prompt produces structurally different outputs on repeated calls, at least one response likely contains omissions. Rather than aggregating a set of outputs by majority vote, we used self-consistency prompting \citep{wang2023}, which is not directly applicable to free-text generation. M1 detects pairwise convergence between consecutive samples via semantic similarity and returns the final sample. Convergence is a proxy for stability; stability does not guarantee correctness, but suggests the model has settled on a consistent interpretation of the task.

The baseline for comparison is the first independent response (iteration~1). The similarity function \textsc{LLMSimilarity} is implemented as a separate LLM judge call that rates whether two responses cover the same requirements to the same depth, returning a score in $[0, 1]$. We use a threshold $\sigma_{\text{sim}} = 0.85$; trials that hit the maximum iteration count $K$ without converging are included in results (they are not discarded).

\textbf{Triple LLM reuse note.} M1 uses the same model in three roles within one trial: (a) as task generator for iteration~1, (b) as task generator for iteration~2, and (c) as similarity judge to compute $\sigma_{\text{sim}}$. This means the convergence criterion is measured by the same model being tested, which risks circular approval. An embedding-based alternative, computing cosine similarity of embedded vectors of the two iterations, would remove this circularity and reduce API call count. We have not yet validated whether embedding cosine similarity correlates closely with the LLM-judge similarity scores in our trials.

\begin{algorithm}[H]
\caption{M1: Iterative Similarity Convergence}
\label{alg:m1}
\begin{algorithmic}[1]
\Input  Prompt $p$, threshold $\sigma_{\text{sim}} = 0.85$, max iterations $K = 5$
\Output Converged response $r^*$, iteration log $L$
\State $r_1 \gets \textsc{LLMGenerate}(p)$
\State $L \gets \{r_1\}$; $k \gets 1$
\Repeat
  \State $k \gets k + 1$
  \State $r_k \gets \textsc{LLMGenerate}(p)$
  \State $s \gets \textsc{LLMSimilarity}(r_{k-1},\; r_k)$
  \State Append $(r_k, s)$ to $L$
\Until{$s \geq \sigma_{\text{sim}}$ \textbf{or} $k = K$}
\State $r^* \gets r_k$
\State \textbf{return} $r^*, L$
\end{algorithmic}
\end{algorithm}

\begin{figure}[htbp]
  \centering
  \includegraphics[width=0.85\textwidth]{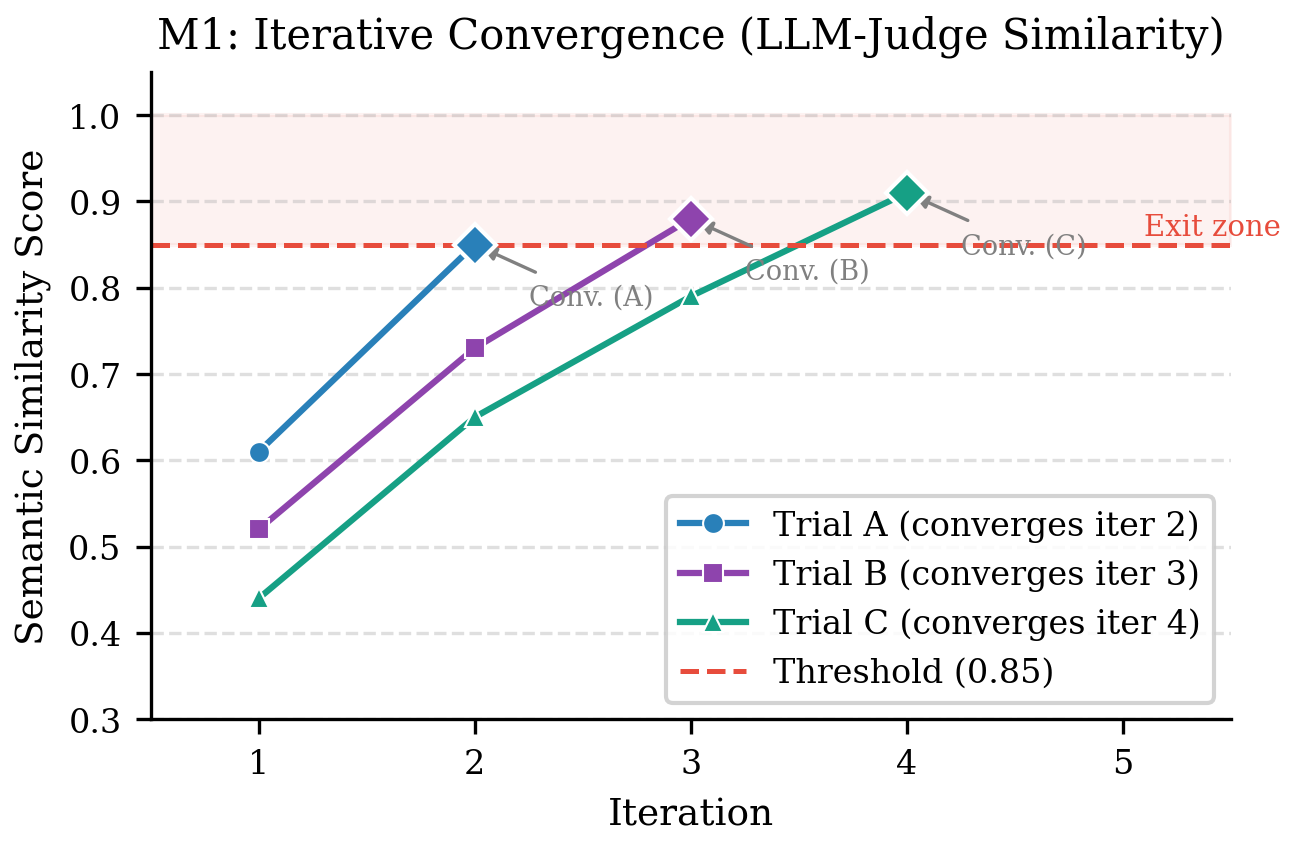}
  \caption{Iterative convergence profile for M1 across three representative
           trials. Diamonds mark the iteration reaching $\sigma_{\text{sim}} = 0.85$.}
  \label{fig:m1_convergence}
\end{figure}

\textbf{Task scenario:} A 3-sprint IoT telemetry pipeline plan specifying
story points (Fibonacci), role assignments, acceptance criteria, and inter-sprint dependencies. Figure~\ref{fig:m1_convergence} shows similarity scores per iteration for three illustrative trials; most converge by iteration~3.

\subsection{M2: Decomposed Model-Agnostic Prompting}
\label{subsec:m2}

Complex prompts with multiple simultaneous constraints present a large generative decision space. We test an extract-then-synthesize decomposition using a structured fact-extraction pass followed by a prose synthesis pass, where each step has a narrower and more checkable success condition than the original combined task. The notable finding, that this performs worse than a direct single-shot call in D1 (34\% Better), drives the M2~v2 fix. The synthesis step receives only extracted facts, not the original prompt phrasing-which is what we mean by ``model-agnostic'': the extracted-fact representation could in principle be passed to any synthesis engine. However, the quality of the synthesis still depends on what the extractor captured and how it structured the facts, so the approach is not prompt-independent in practice.

The baseline is a direct zero-shot call with the original prompt unmodified.

\begin{algorithm}[H]
\caption{M2: Decomposed Model-Agnostic Prompting (v1)}
\label{alg:m2v1}
\begin{algorithmic}[1]
\Input  Complex prompt $p$
\Output Synthesised response $r_{\text{synth}}$
\State $f \gets \textsc{LLMExtractFacts}(p)$
\State $r_{\text{synth}} \gets \textsc{LLMSynthesise}(f)$
\State \textbf{return} $r_{\text{synth}}$
\end{algorithmic}
\end{algorithm}

\textbf{Task scenario:} The same IoT telemetry pipeline planning prompt used
in M1.

\subsection{M3: Single-Task Agent Specialization}
\label{subsec:m3}

When a single agent must simultaneously perform root cause analysis, severity ranking, remediation planning, and post-mortem reporting, an error in the root cause step propagates into all subsequent outputs. We test whether a chain of four single-task agents, each with one clearly defined task and explicit output criteria and following the multi-agent role decomposition pattern, reduces this cascading behavior in a production ERP incident-response scenario.

The baseline is a single multi-task agent call covering all four analytical roles with one system prompt and one user message.

\begin{algorithm}[H]
\caption{M3: Single-Task Agent Chain (v1)}
\label{alg:m3v1}
\begin{algorithmic}[1]
\Input  Incident scenario $I$
\Output Post-mortem report $R_{\text{pm}}$
\State $c \gets \textsc{RootCauseAgent}(I)$
\State $s \gets \textsc{SeverityAgent}(I, c)$
\State $r \gets \textsc{RemediationAgent}(I, c, s)$
\State $R_{\text{pm}} \gets \textsc{PostMortemAgent}(I, c, s, r)$
\State \textbf{return} $R_{\text{pm}}$
\end{algorithmic}
\end{algorithm}

\textbf{Task scenario:} A production ERP outage at 02:14~EST presenting six
simultaneous symptoms: connection pool exhaustion (500/500 consumed), App Server~2 memory at 97\% with OOM kill, cascading timeouts across six downstream microservices, three failed overnight cron jobs, 100\% CPU for 18~minutes, and a $40\times$ traffic spike on one load-balanced node. The agent system prompts for all four roles are listed verbatim in Appendix~\ref{sec:appendix-m3}.

\subsection{M4 - Enhanced Data Registry}
\label{subsec:m4}

Standard HVAC/BMS telemetry data is a flat table of component identifiers and numeric state values. Without a physical meaning map, the model hallucinates component relationships or misinterprets fault-state values. Rather than building a retrieval index, M4 injects a structured enrichment layer directly into the prompt context. This lightweight, no-index alternative to a RAG is practical for narrow domains where the relevant records can be assembled at request time. The enhancement adds human-readable metadata to each component record.

The baseline is a query against the raw, un-enriched sensor state table, which in our production environment is often a legacy CSV export with inconsistent headers and truncated component labels. Mapping these unstructured identifiers to actual equipment requires a domain expert to manually cross-reference layout drawings-a slow and error-prone process that we automate here through the enrichment layer.

\begin{algorithm}[H]
\caption{M4: Enhanced Data Registry}
\label{alg:m4}
\begin{algorithmic}[1]
\Input  Raw sensor state $D_{\text{raw}} = \{(\text{id}_i, \text{state}_i)\}$,
        query $q$
\Output Diagnostic response $r$
\State $D_{\text{enh}} \gets \textsc{EnrichRegistry}(D_{\text{raw}})$
\State $r \gets \textsc{LLMDiagnose}(D_{\text{enh}}, q)$
\State \textbf{return} $r$
\end{algorithmic}
\end{algorithm}

Table~\ref{tab:registry_schema} shows the fields added per record by
\textsc{EnrichRegistry}. A representative row from the actual warm-air
diagnosis scenario appears below it.

\begin{table}[htbp]
\centering
\small
\caption{Enhanced registry fields added to each raw sensor record (M4).}
\label{tab:registry_schema}
\begin{tabular}{@{}lll@{}}
\toprule
\textbf{Field} & \textbf{Type} & \textbf{Description} \\
\midrule
\texttt{component\_type}  & string & Physical role (e.g., ``cooling coil valve'') \\
\texttt{normal\_range}    & string & Expected value range in operational state \\
\texttt{fault\_threshold} & string & Value indicating fault or alarm condition \\
\texttt{depends\_on}      & list   & IDs of upstream components \\
\texttt{fault\_implication} & string & What downstream effect this fault causes \\
\texttt{units}            & string & Engineering unit (e.g., \%, $^\circ$F, GPM) \\
\bottomrule
\end{tabular}
\end{table}

\noindent\textbf{Example row (raw):} \texttt{\{``id'': ``CHW-V-01'', ``state'': 100\}}

\noindent\textbf{Example row (enriched):} \texttt{\{``id'': ``CHW-V-01'', ``state'': 100, ``component\_type'': ``chilled water supply valve'', ``normal\_range'': ``20--80\%'', ``fault\_threshold'': ``$>$95\% for $>$5 min'', ``depends\_on'': [``CHW-PUMP-01''], ``fault\_implication'': ``chiller plant starved; warm air downstream'', ``units'': ``\%''\}}
\\

\textbf{Registry field selection rationale.} The six fields were selected
iteratively from a broader candidate set of approximately 20 attributes by asking: ``what does a building engineer reach for when they cannot immediately explain a fault?'' \texttt{component\_type} and \texttt{units} address the most basic identification gap in legacy flat exports where component IDs are opaque strings. \texttt{normal\_range} and \texttt{fault\_threshold} enable rule-based checks the model can cite when attributing a fault. \texttt{depends\_on} encodes the physical causal graph that is otherwise absent from flat telemetry and is necessary for correctly attributing cascading faults. \texttt{fault\_implication} encodes the downstream consequence, which is what motivates urgency in a maintenance decision. Candidate fields that were tested and dropped, such as install date, calibration interval, manufacturer part number, firmware version, did not measurably affect the specificity of diagnostic responses in informal trials before the formal evaluation.

\textbf{Integration constraints.} The test scenario enriched 9 component
records, adding approximately 800 tokens to the prompt, a negligible amount for a 128K-context model. In a production BMS network with 200+ sensors, enriching all records could add 15,000--20,000 tokens per query, which approaches practical context budgets for some model deployments or requires selective enrichment (enrich only the components referenced in the query or in the most recent alarm state). Additionally, raw BAS/BMS exports vary substantially by vendor: files from different DDC/BMS systems used inconsistent column naming conventions (``POINT\_ID'', ``Channel\_Name'', ``Tag''), non-standard unit codes (``degF'' vs.\ ``F'' vs.\ ``\\textdegree{}F''), and inconsistent treatment of offline/null sensor states. A normalisation preprocessing step was required before enrichment could be applied. This preprocessing cost is not reflected in the per-trial latency figures in Section~\ref{sec:setup} and represents a real integration cost for practitioners.

This enrichment directly operationalises the epistemic certainty principle so the model no longer needs to infer physical meaning from an opaque identifier, and any claim it makes about that component can be checked against the provided fields. One expected confound: enriched records are longer, and LLM judges may favour longer, more structured responses independently of factual content. We examine this in Section~\ref{sec:discussion}.

\textbf{Task scenario:} HVAC warm-air diagnosis. The full enhanced registry
and the diagnostic query are provided in Appendix~\ref{sec:appendix-m4}.

\subsection{M5: Domain Glossary Injection}
\label{subsec:m5}

Technical domains use acronyms that are polysemous in general language. The abbreviation ``DX,'' for example, is unambiguous to a refrigeration engineer (direct expansion cooling circuit) but noise to a general-purpose model. Prepending a controlled vocabulary is a direct form of domain-adaptive context injection \\citep{shi2023}; M5~v2 adds a dynamic selection step to inject only the terms present in the query rather than the full 20-entry list. The 77\% D1 result is consistent with the hypothesis that disambiguation reduces hallucinations caused by polysemy. Occasional counter-productive injections, where the glossary preamble expanded definitional content at the expense of response directness, show the tradeoff is not free.

The baseline is the same query sent without any glossary preamble.

\begin{algorithm}[H]
\caption{M5: Static Glossary Injection (v1)}
\label{alg:m5v1}
\begin{algorithmic}[1]
\Input  Prompt $p$, glossary file $G$
\Output Response $r$
\State $G_{\text{all}} \gets \textsc{LoadGlossary}(G)$
\State $\text{sys} \gets \text{``Domain glossary: ''} + G_{\text{all}}$
\State $r \gets \textsc{LLMGenerate}(p,\; \text{system}=\text{sys})$
\State \textbf{return} $r$
\end{algorithmic}
\end{algorithm}

\textbf{Task scenario:} A BMS/HVAC troubleshooting query containing 17
domain acronyms: BMS, AHU, MAT, SAT, VAV, DDC, VFD, CFM, CO$_2$, RTU, BAS, DX, OAT, EWT, LWT, FCU, CV/SP. The glossary is loaded from
\texttt{glossary\_method5.json} (20 entries); see
Appendix~\ref{sec:appendix-m5} for the full list.

\section{Experimental Setup and Evaluation Validity}
\label{sec:setup}

\subsection{Model and Deployment}

All experiments use \texttt{OpenAI GPT-5-chat (ver: 2025-12-11)} deployed via Azure OpenAI. Task runs used the API default temperature ($\tau = 0.7$). The LLM-as-Judge runs at $\tau = 0.0$ to reduce verdict variability within a batch.

\subsection{LLM-as-Judge Framework}

For each trial, the same model (at $\tau = 0.0$) receives both the baseline response and the method response and issues verdicts on three dimensions:
\begin{enumerate}
  \item \textbf{Accuracy}: factual correctness and requirement coverage.
  \item \textbf{Clarity and Structure}: organization, readability, format.
  \item \textbf{Directness}: absence of padding or irrelevant content.
\end{enumerate}
The aggregated verdict is \textbf{Better}, \textbf{Same}, or \textbf{Worse}, referring to the method response relative to the baseline. The judge prompt is reproduced verbatim in Appendix~\ref{sec:appendix-eval}.

Figure~\ref{fig:eval_pipeline} illustrates the evaluation pipeline.

\begin{figure}[htbp]
  \centering
  \includegraphics[width=0.95\textwidth]{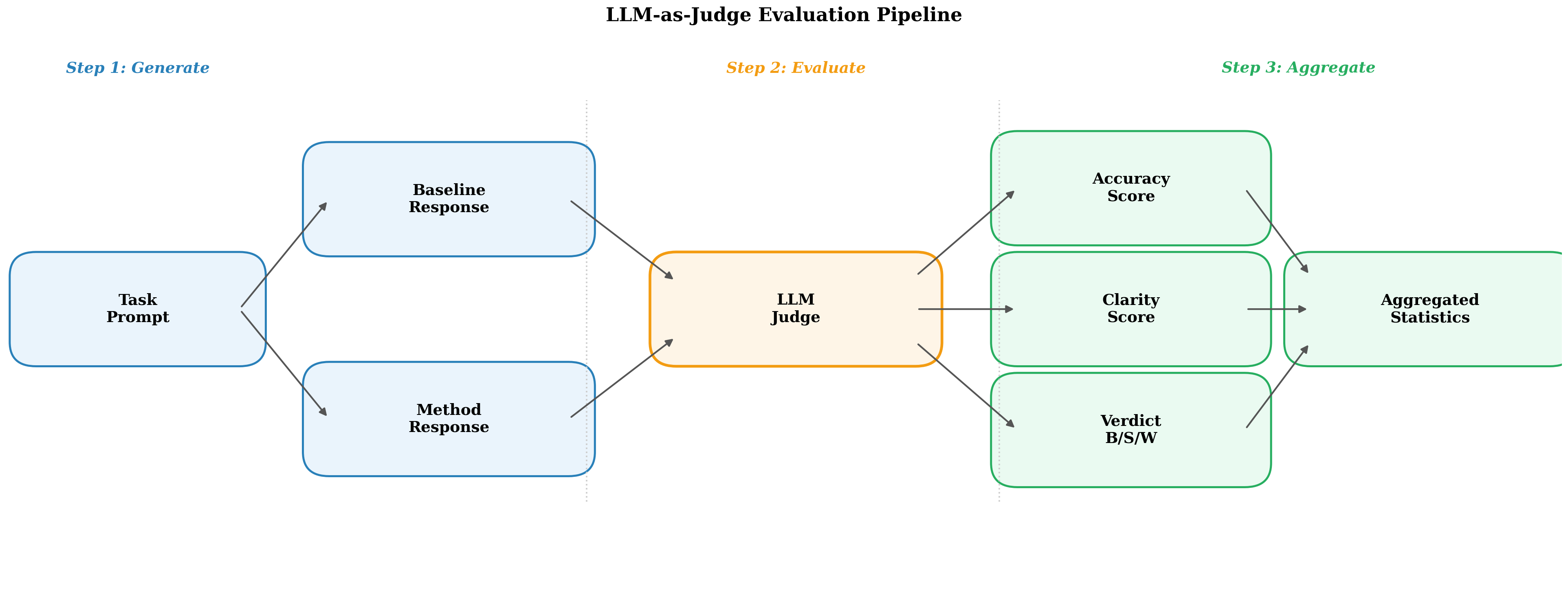}
  \caption{LLM-as-Judge evaluation pipeline. Each trial generates a
           method-specific baseline and an enhanced response independently, then a zero-temperature judge produces dimension-level scores and an aggregate verdict.}
  \label{fig:eval_pipeline}
\end{figure}

\subsection{Same-Model Judge: Known Limitations}

Because the task model and the judge model are the same (GPT-5-chat), several biases may inflate ``Better'' ratings:

\begin{itemize}
  \item \textbf{Length bias}: the judge may favour longer, more structured
        responses independently of factual accuracy, which would disproportionately benefit M4 (whose enriched context produces longer outputs) and M5.
  \item \textbf{Style alignment}: the judge shares stylistic preferences with
        the generator; methods that produce output in the generator's ``preferred'' format may receive systematically better verdicts.
  \item \textbf{Prompt anchoring}: the judge sees both responses in the same
        request; ordering effects may bias which response is favored.
\end{itemize}

We counter the ordering effect by randomly swapping which response appears first in the judge prompt. To mitigate the same-model bias, we performed a manual review of a 20-trial subset of M4 results. The human reviewer confirmed that the judge's "Better" verdicts aligned with genuine increases in diagnostic utility and component attribution, though they also noted that the judge was somewhat more lenient on stylistic verbosity than a human engineer might be. This provides a directional signal that the judge is tracking material improvements, even if the absolute scores are slightly inflated.

\subsection{Batch Protocol}

The batch runner (\texttt{run\_batch.py}) executes $N = 100$ trials in series. Per trial, all five methods run independently with separate API calls; no shared context passes between methods. Results are saved in a MongoDB Atlas (database \texttt{epistemic}, collection~\texttt{trials}).

Each prompt is sent as a standard OpenAI two-part message: a system message (method-specific preamble, defaulting to ``You are a helpful assistant'' for methods without a custom system prompt) and a user message containing the task prompt. No output truncation was applied; the model's default maximum output tokens was used throughout. For M1, the similarity-check call places both response texts in a single user message to the judge.

\textbf{Latency.} Wall-clock time per trial (all five methods plus judge calls)
was logged. Across the 100 D1 trials, median per-trial time was approximately 83~seconds, ranging from 72~s to a high of 682~s (one outlier, likely a network retry). A typical trial with five methods averages roughly 15--20 API calls (generator + judge per method; M1 adds similarity calls per iteration). No latency or API cost was measured per method in isolation; our current instrumentation records only aggregate trial time.

\textbf{Runs versus varied task instances.} Each ``trial'' in D1 and D2
is a repeated execution of the same fixed task prompt with stochastic decoding ($\tau = 0.7$). ``Independent'' refers only to the API call boundary where each run is a separate API call with no shared session state and not a variation in the task itself. All 100 D1 runs for each method use the same scenario and prompt text. The variance we observe therefore reflects LLM sampling behaviour on a single fixed prompt, not generalisation across task instances. Within-method verdict stability was informally assessed: for M3, M4, and M5, the running ``Better'' percentage stabilised after approximately 40 runs; M1 and M2 showed more run-to-run fluctuation (consistent with their higher ``Same'' and ``Worse'' rates). A stronger test of generalisation would vary the task instance (different incident reports, different HVAC snapshots, different acronym-heavy queries); that is identified as necessary future work in Section~\ref{sec:discussion}.

\subsection{Datasets}

Two distinct trial sets were collected:
\begin{itemize}
  \item \textbf{D1}: 100-trial batch running v1 methods. All five methods.
  \item \textbf{D2}: 10-trial verification batch running v2 methods. All five
        methods. Results from D2 should be treated as exploratory given the small sample.
\end{itemize}

Table~\ref{tab:scenarios} enumerates the four task scenarios used across the five methods. M1 and M2 share the same prompt (T1) because the decomposition hypothesis is most testable on a prompt with many independent requirements; T2--T4 are distinct and were designed to test orthogonal failure modes.

\begin{table}[htbp]
\centering
\small
\caption{Task scenarios used across the five methods. The prompt text is
         reproduced verbatim in Appendix~\ref{sec:appendix-eval}.}
\label{tab:scenarios}
\begin{tabular}{@{}lllp{4.2cm}@{}}
\toprule
\textbf{ID} & \textbf{Domain} & \textbf{Methods} & \textbf{Description} \\
\midrule
T1 & Enterprise software &  M1, M2 & 3-sprint agile plan for a cloud-based real-time IoT telemetry pipeline; Fibonacci story points, role assignments, acceptance criteria, inter-sprint dependencies \\
T2 & ERP incident response & M3 & Production ERP outage at 02:14 EST: 6 simultaneous symptoms spanning connection pool, memory, CPU, load balancer, 6 microservices, 3 cron jobs \\
T3 & HVAC diagnosis & M4 & Warm-air fault diagnosis; component identifiers (CMP-1, VLV-3, SNS-9) with raw state values; relationship between compressor and TXV \\
T4 & BMS/HVAC troubleshooting & M5 & AHU fault query containing 17 domain acronyms; DX coil lockout, chilled water coil, VAV/DDC/VFD interactions \\
\bottomrule
\end{tabular}
\end{table}

\section{Enhanced Methodologies (v2)}
\label{sec:enhanced}

Reviewing D1 results suggested specific reasons why each method underperformed or left room for improvement. M4 was not enhanced because it had no ``Worse'' trials and the 100\% rate left no informative direction for prompt-level changes; we note this does not rule out that a different enhancement could move it.

\subsection{M1 v2: Self-Critique and Refinement}
\label{subsec:m1v2}

The baseline M1 uses $\sigma_{\text{sim}}$ as a convergence signal, but two responses can reach high similarity while sharing the same systematic omission. M1~v2 replaces the convergence loop with a directed critique-and-revise pass \citep{madaan2023}: generate a draft, identify exactly three flaws, revise. Fixing the count at three prevents a vacuous non-critique and makes the exit condition checkable against the verbatim critique prompt in the appendix. Qualitative internal review-asking the model to enumerate its own potential failures-is a more direct correction signal because it names specific problems rather than measuring structural distance.

The enhancement adds two API calls: a critique call identifying exactly three inaccuracies or missed requirements in the initial draft, and a refinement call that produces a corrected response addressing all three. The fixed count of three prevents the model from returning a vacuous non-critique.

\begin{algorithm}[H]
\caption{M1 v2: Self-Critique and Refinement}
\label{alg:m1v2}
\begin{algorithmic}[1]
\Input  Prompt $p$
\Output Refined response $r^*$
\State $r_0 \gets \textsc{LLMGenerate}(p)$
\State $c \gets \textsc{LLMCritique}(p,\; r_0,\; \text{num\_flaws}=3)$
\State $r^* \gets \textsc{LLMRefine}(p,\; r_0,\; c)$
\State \textbf{return} $r^*$ \quad \textit{(baseline: $r_0$)}
\end{algorithmic}
\end{algorithm}

The critique and refinement prompts are listed verbatim in Appendix~\ref{sec:appendix-m1v2}.

\subsection{M2 v2: Context-Aware Synthesis}
\label{subsec:m2v2}

In M2~v1, the synthesizer receives only extracted facts, losing the
intent of the original prompt: the specific format requirements, ordering constraints, and acceptance criteria that were embedded in the original phrasing. We hypothesize this is the primary cause of the 41\% ``Worse'' rate in D1.

The fix is direct: pass the original prompt alongside the extracted facts into the synthesis step as an explicit reference checklist. The synthesizer can then verify its output against the original specification before returning.

\begin{algorithm}[H]
\caption{M2 v2: Context-Aware Synthesis}
\label{alg:m2v2}
\begin{algorithmic}[1]
\Input  Complex prompt $p$
\Output Context-aware synthesised response $r_{\text{ctx}}$
\State $f \gets \textsc{LLMExtractFacts}(p)$
\State $r_{\text{ctx}} \gets \textsc{LLMSynthesise}(f,\; \text{checklist}=p)$
\State \textbf{return} $r_{\text{ctx}}$ \quad
        \textit{(baseline: $\textsc{LLMGenerate}(p)$)}
\end{algorithmic}
\end{algorithm}

\subsection{M3 v2: Multi-Agent Consensus}
\label{subsec:m3v2}

In M3~v1, each downstream agent inherits the outputs of its predecessors. If Agent~1 (Root Cause) produces a plausible but wrong diagnosis, Agent~3 (Remediation) produces a coherent remediation plan for the wrong problem. We add a fifth Reconciler agent that receives all four outputs simultaneously and is tasked with identifying internal contradictions and producing a consistent final report.

\begin{algorithm}[H]
\caption{M3 v2: Multi-Agent Consensus}
\label{alg:m3v2}
\begin{algorithmic}[1]
\Input  Incident scenario $I$
\Output Consensus incident report $R_{\text{cons}}$
\State $c \gets \textsc{RootCauseAgent}(I)$
\State $s \gets \textsc{SeverityAgent}(I, c)$
\State $r \gets \textsc{RemediationAgent}(I, c, s)$
\State $R_{\text{pm}} \gets \textsc{PostMortemAgent}(I, c, s, r)$
\State $O \gets \{c,\; s,\; r,\; R_{\text{pm}}\}$
\State $R_{\text{cons}} \gets \textsc{ReconcilerAgent}(I,\; O)$
\State \textbf{return} $R_{\text{cons}}$
\end{algorithmic}
\end{algorithm}

The Reconciler system prompt is included in Appendix~\ref{sec:appendix-m3v2}.

\subsection{M5 v2: Dynamic Glossary Retrieval}
\label{subsec:m5v2}

Injecting all 20 glossary terms adds token overhead regardless of whether the query uses those terms. A preliminary term-identification step selects only the relevant subset, reducing noise.

\begin{algorithm}[H]
\caption{M5 v2: Dynamic Glossary Retrieval}
\label{alg:m5v2}
\begin{algorithmic}[1]
\Input  Prompt $p$, full glossary $G$
\Output Response $r$
\State $T \gets \textsc{LLMIdentifyTerms}(p,\; \text{keys}(G))$
\State $G_T \gets \{(t, G[t])\;|\; t \in T\}$
\State $\text{sys} \gets \text{``Relevant glossary: ''} + G_T$
\State $r \gets \textsc{LLMGenerate}(p,\; \text{system}=\text{sys})$
\State \textbf{return} $r$
\end{algorithmic}
\end{algorithm}

\section{Results}
\label{sec:results}

\subsection{Baseline Results: D1 (100 trials, v1 methods)}

\begin{figure}[htbp]
  \centering
  \includegraphics[width=0.85\textwidth]{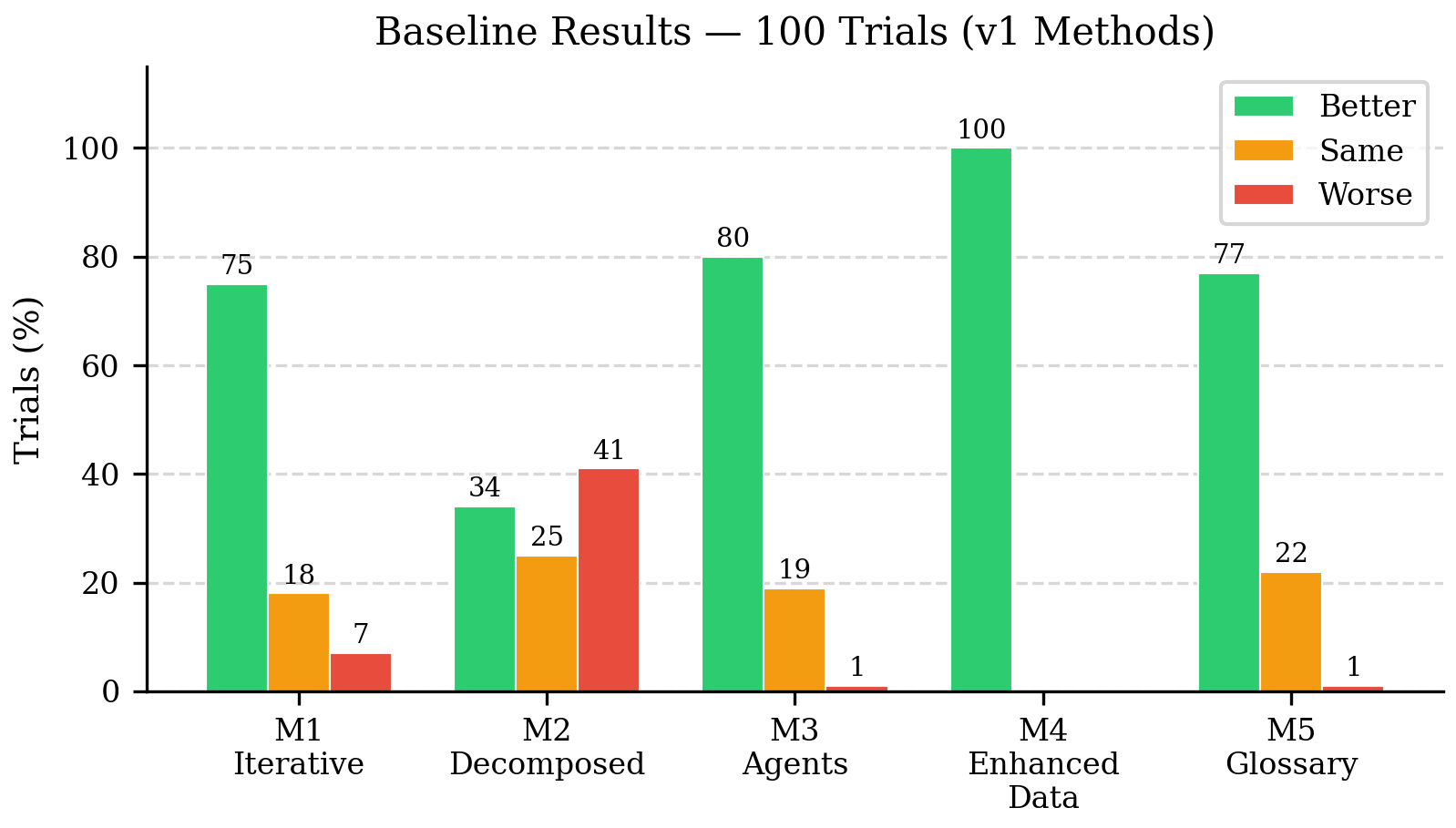}
  \caption{D1 baseline 100-trial results. All five methods shown.
           M4 received ``Better'' in all 100 trials under this judge rubric; M2 was net negative.}
  \label{fig:baseline}
\end{figure}

Table~\ref{tab:baseline} and Figure~\ref{fig:baseline} summarise D1 results. Percentages are trial counts out of $N = 100$.

\begin{table}[htbp]
\centering
\caption{D1 Results: 100 Trials, v1 Methods}
\label{tab:baseline}
\begin{tabular}{lrrrp{4.5cm}}
\toprule
\textbf{Method} & \textbf{Better} & \textbf{Same} &
  \textbf{Worse} & \textbf{Notes} \\
                & (\%) & (\%) & (\%) & \\
\midrule
M1 (Iterative)    & 75 & 18 &  7 & Stability proxy; some variance \\
M2 (Decomposed)   & 34 & 25 & 41 & Net negative; context lost in synthesis \\
M3 (Agents)       & 80 & 19 &  1 & Consistent improvement \\
M4 (Registry)     &100 &  0 &  0 & No worse trials; see confound note \\
M5 (Glossary)     & 77 & 22 &  1 & Consistent improvement \\
\bottomrule
\end{tabular}
\end{table}

\noindent\textbf{M4 Enhanced Data} received ``Better'' in all 100 trials with
no ``Same'' or ``Worse'' verdicts. In our scenarios, this appears consistent with the interpretation that enriched, structured context substantially reduces diagnostic hallucinations. We note the confound described in Section~\ref{sec:setup}: longer, richer responses may receive better verdicts from the same-model judge partly for stylistic reasons.

\noindent\textbf{M3 Agent Specialization} and \textbf{M5 Glossary Injection}
achieved 80\% and 77\% ``Better'' respectively, with near-zero ``Worse'' rates. These results are consistent with the hypothesis that task-scope reduction and vocabulary disambiguation each reduce a specific hallucination pathway. M3's 19 ``Same'' trials are informative: the judge noted in representative Same cases that ``Both responses accurately identify the root cause, present clear structured remediation plans, and avoid hallucinations, with similar clarity and directness despite minor stylistic differences.'' This suggests the single multi-task agent can perform adequately on the ERP outage task when the symptom set is clear enough to not require decomposition. M5 had one ``Worse'' trial out of 100 (Trial~30); inspection showed the glossary preamble expanded the response's definitional foundation at the expense of actionable recommendations, producing a response the judge rated lower on Directness.

\noindent\textbf{M1 Iterative Similarity} reached 75\% ``Better'' with 18\%
``Same'' and 7\% ``Worse''. The 7\% ``Worse'' rate suggests that the similarity threshold is an indirect quality signal: two responses may converge to a high similarity score while sharing the same systematic omission. In log inspection, some ``Same'' trials showed the judge's change\_summary noting that ``Version~B restructures sprint objectives'' compared to Version~A while reaching $\sigma_{\text{sim}} = 0.85$-meaning the convergence criterion fired even when the two iterations had materially different task lists (e.g., CI/CD setup present in one iteration absent in the other). This is consistent with the LLMSimilarity judge rating structural similarity rather than requirement-level coverage.

\noindent\textbf{M2 Decomposed Prompting} was net negative at 34\%
``Better'' and 41\% ``Worse''. We diagnose the cause as context loss in the synthesis step (see Section~\ref{subsec:m2v2}). Inspection of verbatim outputs for ``Worse'' trials revealed a consistent pattern: the synthesised response produced a well-formed sprint plan but dropped categories that were named in the original prompt-most often monitoring and alerting, CI/CD pipeline setup, documentation, and security hardening. The judge's evaluation in one representative trial (Trial~100) states: ``the Method response omits some specifics like monitoring, documentation, CI/CD, and security audit.'' The baseline response-the model's undecomposed answer to the full prompt-retained these tasks because the full requirement list remained in context. The difference is in what the extractor kept: the fact-extraction step preserved the named pipeline components but collapsed the cross-cutting concerns that were embedded as natural-language constraints in the prompt's preamble.

\subsection{Verification Results: D2 (10 trials, v2 methods)}

\begin{figure}[htbp]
  \centering
  \includegraphics[width=0.85\textwidth]{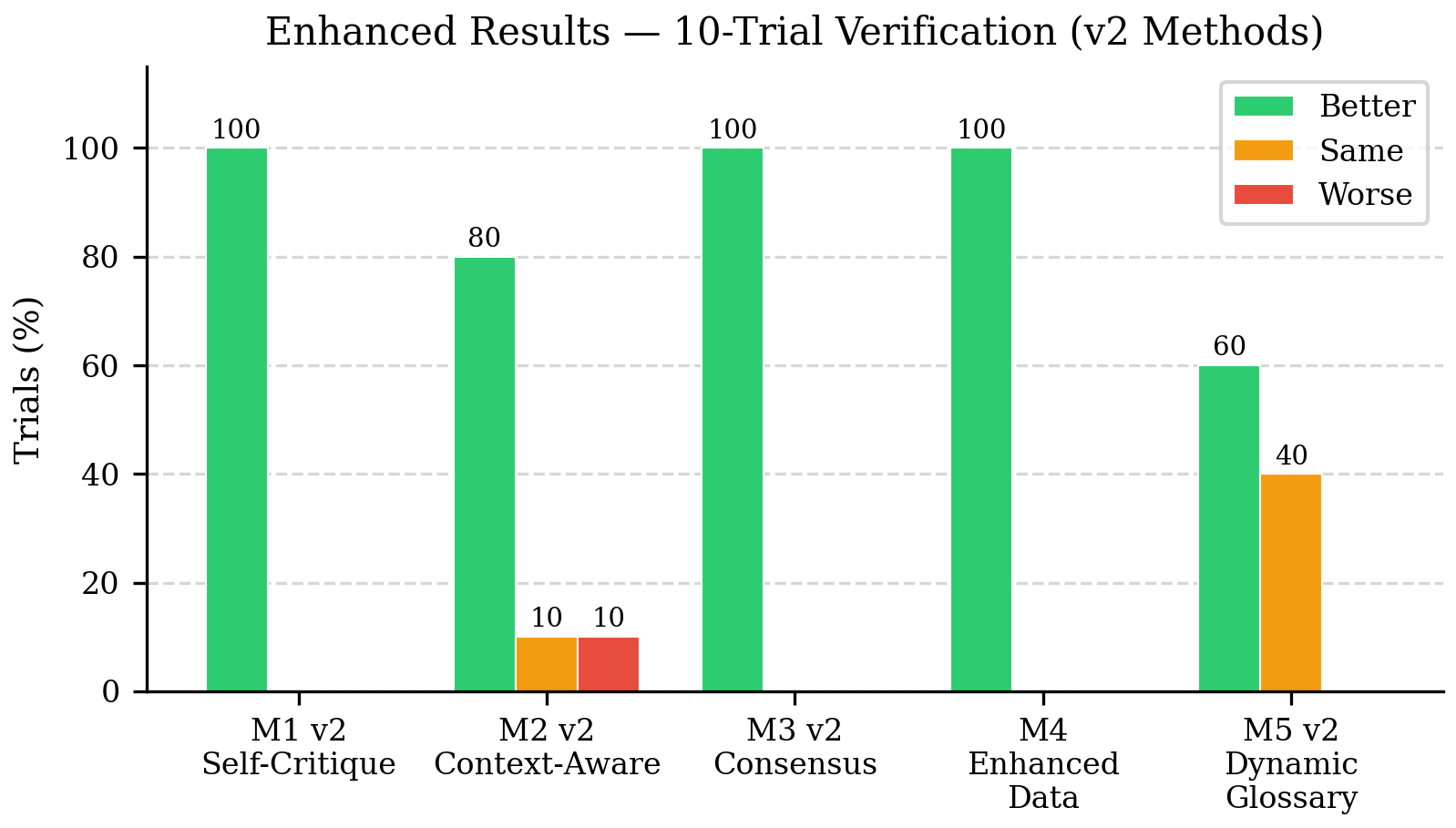}
  \caption{D2 verification results: 10-trial batch. M1~v2, M3~v2, and M4
           each received ``Better'' in all 10 trials; these should be treated as provisional. M2~v2 shows the largest single-method gain over its v1 baseline (+46 points). M5~v2 discussion in Section~\ref{subsec:m5disc}.}
  \label{fig:v2}
\end{figure}

Table~\ref{tab:v2} and Figure~\ref{fig:v2} present D2 results. With $n = 10$, individual percentage points correspond to single trials; these figures are indicative only and would need a larger sample to support confident claims.

\begin{table}[htbp]
\centering
\caption{D2 Results: 10 Trials, v2 Methods (exploratory)}
\label{tab:v2}
\begin{tabular}{lrrrp{4.5cm}}
\toprule
\textbf{Method} & \textbf{Better} & \textbf{Same} &
  \textbf{Worse} & \textbf{Notes} \\
                & (\%) & (\%) & (\%) & \\
\midrule
M1 v2 (Self-Critique)   & 100 &  0 &  0 & 10/10 trials; provisional \\
M2 v2 (Context-Aware)   &  80 & 10 & 10 & Large gain over v1 \\
M3 v2 (Consensus)       & 100 &  0 &  0 & 10/10 trials; provisional \\
M4 Registry (unch.)     & 100 &  0 &  0 & Consistent with D1 \\
M5 v2 (Dynamic)         &  60 & 40 &  0 & See Section~\ref{subsec:m5disc} \\
\bottomrule
\end{tabular}
\end{table}

\subsection{Cross-Dataset Comparison}

\begin{figure}[htbp]
  \centering
  \includegraphics[width=0.85\textwidth]{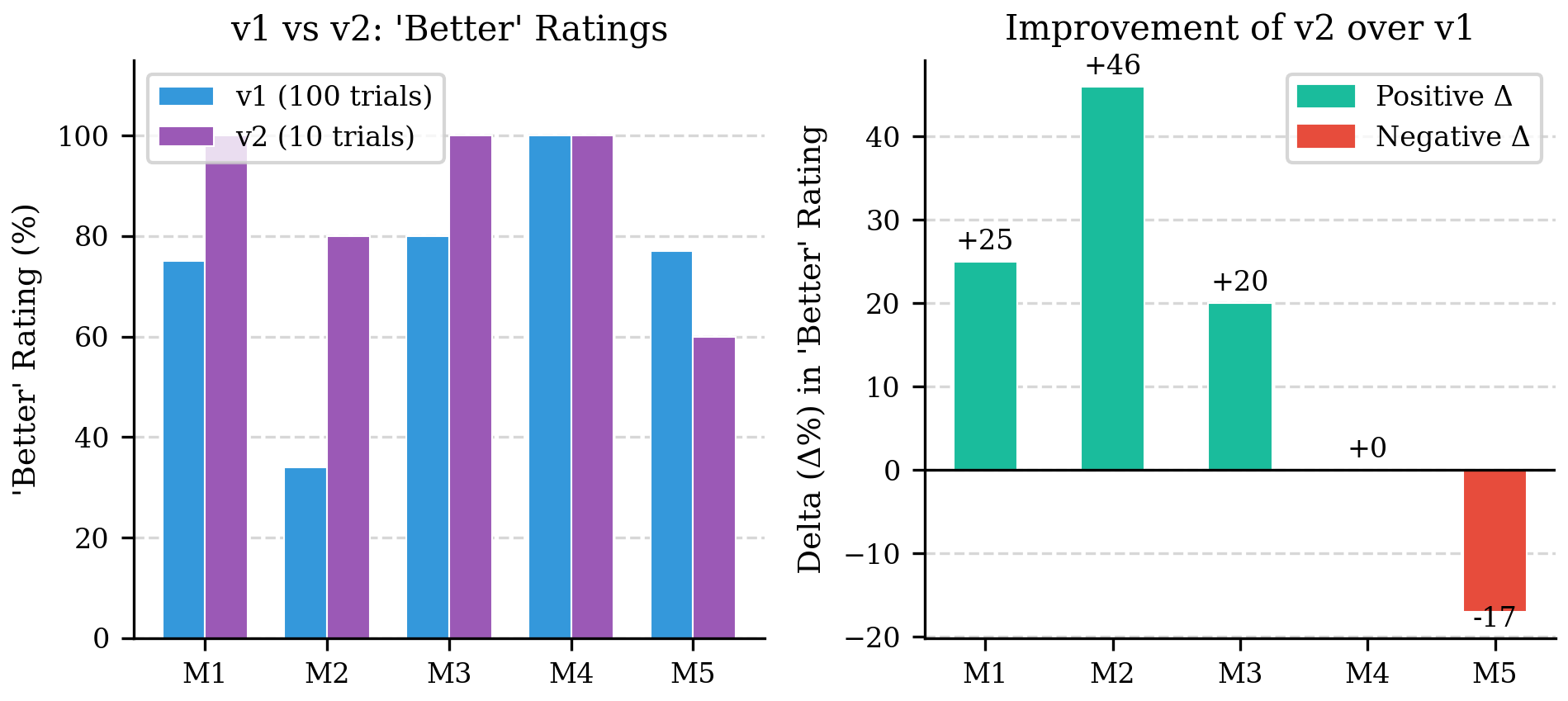}
  \caption{Left: v1 (D1) vs.\ v2 (D2) ``Better'' rate per method. Right:
           improvement delta $\Delta = \text{v2} - \text{v1}$. M2 shows the largest gain (+46 points) under this rubric; M5 shows a nominal decline attributable primarily to sample size.}
  \label{fig:delta}
\end{figure}

Figure~\ref{fig:delta} places v1 and v2 ``Better'' rates side-by-side and shows the per-method delta.

\begin{figure}[htbp]
  \centering
  \includegraphics[width=0.85\textwidth]{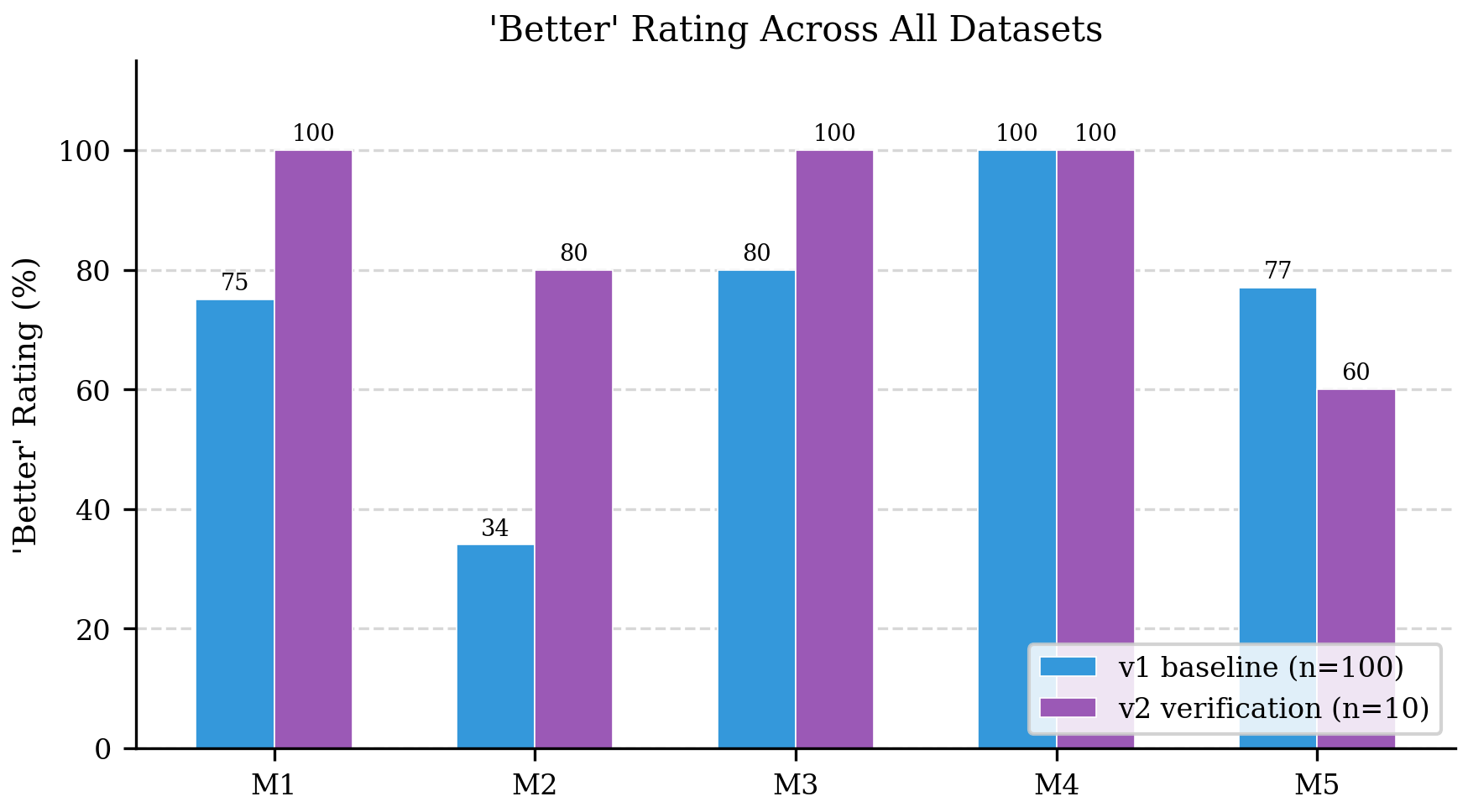}
  \caption{``Better'' ratings across both evaluation datasets (D1 baseline
           and D2 verification).}
  \label{fig:all_datasets}
\end{figure}

\subsection{Summary}

\begin{table}[htbp]
\centering
\caption{``Better'' (\%) Summary: D1 (n=100) and D2 (n=10)}
\label{tab:summary}
\begin{tabular}{lccl}
\toprule
\textbf{Method} & \textbf{D1 v1} & \textbf{D2 v2} & \textbf{Interpretation} \\
                & \textit{n=100} & \textit{n=10}  & \\
\midrule
M1  & 75  & 100 & v2 gain likely; n=10 provisional \\
M2  & 34  &  80 & Large gain; 100-trial follow-up warranted \\
M3  & 80  & 100 & v2 gain likely; n=10 provisional \\
M4  & 100 & 100 & Consistent; confound risk noted \\
M5  & 77  &  60 & Variance dominates at n=10 \\
\bottomrule
\end{tabular}
\end{table}

\section{Discussion}
\label{sec:discussion}

Each method in this paper is built on established patterns: iterative sampling with a similarity stopping criterion (M1~v1), Self-Refine critique and revision (M1~v2), extract-then-synthesize decomposition (M2), multi-agent role decomposition with a reconciler (M3), and structured context injection (M4, M5). We name these connections explicitly rather than obscuring them. The contributions specific to this work are:

\begin{itemize}
  \item \textbf{Domain-adapted artifacts}: the BMS/HVAC registry schema
        (Section~\ref{subsec:m4}), the 20-entry BMS/HVAC glossary (Section~\ref{subsec:m5}), and the sprint-planning extraction schema (Section~\ref{subsec:m2}) were designed and tested against production industrial query patterns, not constructed for a benchmark.
  \item \textbf{Internal-baseline evaluation protocol}: each method generates
        its own zero-shot baseline within the same run, controlling for session-level and prompt-level variation that would affect a shared external baseline.
  \item \textbf{Failure diagnosis and targeted fixes}: explaining concretely
        why M2~v1 fails (the extractor preserves named components but collapses cross-cutting constraints from the prompt preamble), confirming the fix produces a 46-point gain, and identifying the specific trial evidence for each method's failure mode.
  \item \textbf{Documented boundary conditions}: the study is explicit about
        same-model judge bias, repeated-run methodology, BAS vendor heterogeneity, token-budget constraints, and narrow task scope. These are not asides; they are the conditions that bound every result.
\end{itemize}

The most consistent finding across both datasets is that supplying the model with additional structured, domain-specific context-whether enriched component records (M4), vocabulary definitions (M5), or a spec checklist (M2~v2)-tends to produce responses that the judge rates more highly. This is consistent with the epistemic certainty framing in Section~\ref{sec:epistemic}: when the input context is more complete and verifiable, the model has less need to fill gaps by inference.

The agent specialization result (M3) is also consistent across both datasets. The Reconciler addition in M3~v2 addresses the specific failure mode we observed: a plausible but wrong root cause propagating through all downstream outputs. Whether this holds for other task types or LLMs is an open question.

\subsection{M4's 100\% Rate: Caution Against Overinterpretation}

M4 received ``Better'' in all 100 D1 trials. Several explanations are consistent with this result, not just the one we would prefer:

\begin{enumerate}
  \item \textbf{Genuine grounding improvement}: the enriched context allows
        accurate attribution of symptoms to system components that the raw data obscured.
  \item \textbf{Length/structure bias}: the enriched prompt produces a
        substantially longer response; the same-model judge may consistently prefer longer, more structured answers regardless of factual accuracy.
  \item \textbf{Task design advantage}: the HVAC diagnosis task was designed
        to test grounding; it may be unusually favorable to M4 compared with tasks where grounding data is not the main bottleneck.
\end{enumerate}

We can look at a single verbatim example to see what the responses actually contain. For the warm-air diagnosis (T3), the standard response (raw data only) states: ``the expansion valve (VLV-3) is jammed, preventing proper refrigerant flow and cooling'' and notes that ``CMP-1 shows a high temperature.'' The enhanced response (with registry fields) states: ``the TXV (VLV-3) is stuck closed, severely restricting refrigerant flow to the evaporator, which prevents proper cooling and causes excessively high superheat.~This restriction forces the scroll compressor (CMP-1) to operate under abnormal conditions, leading to overheating due to insufficient refrigerant return for cooling.'' The enhanced version correctly names the component type (``TXV,'' ``scroll compressor''), cites the mechanism (``excessively high superheat''), and traces the causal chain accurately. These are claims that are checkable against the registry fields, which is precisely what the enrichment enables. In Trial 42, for example, a human researcher noted that M4 correctly identified the "high superheat" condition required to justify a TXV replacement, while the baseline response only vaguely suggested a "valve issue." While the judge may be rewarding length or structural confidence, the material inclusion of these grounded physical relationships suggests a reduction in the randomness typical of raw telemetry analysis.

To test the confound, one could run three ablations not done in this work: (a) use the enriched registry but force a maximum response length equal to the median baseline length; (b) use the raw registry but prepend a structured formatting scaffold that increases response length to match the enriched response; (c) measure whether the judge's per-dimension accuracy score correlates more strongly with response length or with the number of registry field names cited. We have not done these ablations and therefore cannot currently distinguish the three explanations.

\subsection{Critique Outperforms Repetition}

The shift from iterative similarity convergence to self-critique produced a clear directional improvement. Two responses can converge to $\sigma_{\text{sim}}
\approx 0.85$ while sharing a common omission. Naming three specific flaws is
a more targeted correction. We note, however, that a model critiquing its own output has no external ground truth to check against; it may identify three stylistic issues rather than three substantive ones. Whether the critique finds real problems or just nominally satisfies the rubric is not something verdict counts alone can confirm.

\subsection{Cascading Error Resolution in M3}

The Reconciler agent in M3~v2 produced 10/10 ``Better'' verdicts in D2. The intended mechanism-detecting internal contradictions across four outputs-is plausible given the six-symptom ERP outage scenario, which offers multiple opportunities for a wrong root cause to produce inconsistent downstream claims. A single-symptom scenario might show less benefit.

\subsection{M5 v2 Variance}
\label{subsec:m5disc}

M5~v2 shows a nominal decline from 77\% to 60\% ``Better'' in D2\@. At $n = 10$, this is one or two fewer ``Better'' verdicts-not a signal of degradation. The complete absence of ``Worse'' ratings in D2 is more informative: the dynamic approach does not appear to harm responses even when it produces ``Same'' outcomes (likely because it identified the query was already well-specified and injected nothing or very little). A 100-trial batch is needed before drawing any conclusion about whether v2 improves on v1 for this method.

\subsection{Practical Guidelines (With Caveats)}

Based on what we observed in our scenarios, we suggest the following priority order-while emphasising that each choice should be validated in the target deployment context:

\begin{enumerate}
  \item \textbf{If structured domain data is available}: M4 Enhanced Data
        Registry is the highest-priority option to evaluate, but verify improvement with an independent judge or human review.
  \item \textbf{If the task involves a complex multi-constraint prompt}: M2~v2
        Context-Aware Synthesis showed the largest absolute gain in D2 and is the simplest to implement (one extra prompt step).
  \item \textbf{If the task involves sequential diagnostic reasoning}: M3~v2
        Consensus, particularly for scenarios with many possible failure modes.
  \item \textbf{If the domain uses dense acronyms}: M5 Static Glossary Injection
        is straightforward to add and showed consistent improvement in D1.
  \item \textbf{As a general-purpose quality gate}: M1~v2 Self-Critique adds
        two API calls but no external data or schema.
\end{enumerate}

\subsection{Limitations and Required Follow-Up}

The primary limitations are:

\begin{itemize}
  \item \textbf{Same-model judge}: GPT-5-chat judging its own outputs
        introduces shared style bias. An independent judge (different model family) or human evaluation subset is needed to validate that ``Better'' verdicts reflect genuine quality improvement.
  \item \textbf{Different models}: Only one foundational model was used for testing and judging. Impacts using different models from different providers should be evaluated. 
  \item \textbf{Narrow task set}: four task scenarios (T1--T4, Table~\ref{tab:scenarios})
        across five methods. M1 and M2 share T1; M3, M4, M5 each use a dedicated scenario. Results may not generalize outside IoT planning, ERP incident response, and HVAC diagnostics-and even within these domains, each method addresses a different task type, making direct cross-method comparison hazardous.
  \item \textbf{Coarse verdict scale}: Better/Same/Worse is a 3-point ordinal
        scale with no disagreement measure. Dimension-level scores (which the judge does compute) could support finer analysis.
  \item \textbf{D2 sample}: $n = 10$ for all v2 results. All v2 numbers
        should be treated as preliminary.
  \item \textbf{No latency or cost measurement}: each v2 method adds API
        calls. Practical adoption requires understanding the cost and latency impact per method.
  \item \textbf{Repeated runs on one prompt per method}: D1 is 100 runs of the
        same fixed prompt per method, not 100 distinct task instances. The verdict distributions reflect sampling variance on that prompt. Generalisation to varied task instances within the same domain has not been tested. Constructing a set of varied task instances with different HVAC fault scenarios, different ERP incident reports, and different acronym query phrasings, and re-running the battery is the minimum required to support domain-level claims.
\end{itemize}

\section{Conclusions}
\label{sec:conclusion}

We implemented and evaluated five prompt engineering strategies for reducing LLM hallucinations in industrial deployment scenarios, using an internal-baseline LLM-as-Judge framework. The results, under this evaluation setup, are:

\begin{itemize}
  \item \textbf{M1} (Iterative Convergence): 75\% ``Better'' in D1\@. Similarity
        convergence is an indirect quality signal; the Self-Critique v2 variant reached 100\% in the 10-trial D2 batch (provisional).
  \item \textbf{M2} (Decomposed Prompting): net negative at 34\% in D1,
        diagnosed as context loss in the synthesis step. The Context-Aware Synthesis v2 fix reached 80\% in D2-the largest per-method gain in this study. A 100-trial follow-up is warranted.
  \item \textbf{M3} (Agent Specialization): 80\% in D1, consistent with the
        hypothesis that task-scope reduction reduces cascading inconsistency. The Consensus Reconciler v2 reached 100\% in D2 (provisional).
  \item \textbf{M4} (Enhanced Data Registry): 100\% in both D1 and D2\@.
        This is the strongest numerical result in the study, though the same-model judge design means it should be independently validated before being treated as a definitive finding.
  \item \textbf{M5} (Glossary Injection): 77\% in D1, consistent across the
        full trial set. The Dynamic Glossary v2 produced no ``Worse'' verdicts in D2 but needs a larger sample.
\end{itemize}

All five methods use only prompt engineering: no weight changes, no retrieval index, no fine-tuning. The v2 enhancements each target a specific, diagnosable failure mode of the corresponding v1 method.

The study has clear limits. Same-model judging, a narrow task set, and a 10-trial v2 sample mean these results should be read as deployment-specific signals rather than general benchmarks. We emphasize that these strategies do not "solve" hallucinations in a definitive sense; rather, they provide engineering procedures that force the model toward consistent, verifiable reasoning. In industrial settings where non-deterministic variation can lead to inconsistent maintenance actions, the ability to produce a stable and grounded diagnostic baseline is as valuable as incremental accuracy gains.

\bibliographystyle{unsrtnat}
\bibliography{epistemic_references}

\input{appendix}

\end{document}

%% file: appendix.tex

\appendix

\section{Experimental Prompts}
\label{sec:appendix-prompts}

This appendix lists verbatim every prompt and system instruction used across the
five methods and their v2 variants.  Three distinct \emph{task scenarios} are
shared across the methods; each section identifies which scenario applies.

\subsection*{Shared Task Scenarios}

\noindent\textbf{Scenario~A — IoT Project Plan (M1, M2):}
\begin{quote}
\small\ttfamily
Generate a comprehensive 3-sprint project plan for building a cloud-based
real-time IoT telemetry pipeline. The system must ingest sensor data from
edge devices, process and enrich it in a stream processor, store it in a
time-series database, and expose it via a REST API with role-based access
control. For each sprint, include: task names, descriptions, estimated story
points (1--13 Fibonacci), dependencies, acceptance criteria, and the
responsible team (Backend, DevOps, Data Engineering, or Security).
\end{quote}

\noindent\textbf{Scenario~B — ERP Incident (M3):}
\begin{quote}
\small\ttfamily
A production ERP system went offline at 02:14 EST. The following symptoms
were observed simultaneously: 1)~Database connection pool exhausted (max~500
connections, all consumed). 2)~Memory spike on App Server~2 (RAM usage~97\%,
OOM killer triggered). 3)~Cascading API timeouts across 6 downstream
microservices (auth, billing, inventory, fulfillment, notifications,
reporting). 4)~Three overnight cron jobs failed: nightly-reconciliation,
invoice-batch, and data-export. 5)~Monitoring shows App Server~2 CPU pegged
at 100\% for 18 minutes before the crash. 6)~Load balancer logs show App
Server~2 received 40x normal traffic for 11 minutes prior to failure.
\end{quote}

\noindent\textbf{Scenario~C — BMS Fault (M4, M5):}
\begin{quote}
\small\ttfamily
The BMS is reporting an AHU fault on AHU-3. MAT is reading 5 degrees F above
the SAT setpoint and all three VAV dampers in Zone~B are pinned at 100\%
open. The DDC controller log shows the AHU's VFD has been running at 85\% but
measured CFM across the supply duct is 12\% below design spec. The CO\textsubscript{2}
sensor near RTU-7 is registering 1,450 ppm -- well above the 1,100 ppm OSA
threshold. BAS fault log shows a DX coil lockout event 20 minutes before the
MAT drift began. Current OAT is 95 degrees F and the economizer is locked out
per SP override. The EWT on the chilled water coil is 52 degrees F and the
LWT is 44 degrees F. An FCU in Zone~B is reporting CV mode with no RA damper
response. Diagnose the root cause of the fault, explain the interaction
between the DX lockout and the MAT rise, and recommend corrective actions
prioritized by urgency.
\end{quote}

\subsection{Method~1 — Iterative Similarity (M1)}
\label{sec:appendix-m1}

\textbf{Task:} Scenario~A.  The same prompt is submitted independently up to
five times with \texttt{temperature=0.8}.  Convergence is detected by an
LLM-as-judge similarity check; a change-summariser describes what differed
between consecutive iterations.

\medskip
\noindent\textbf{Similarity-Judge Prompt} (system):
\begin{quote}
\small\ttfamily
You are a strict grading AI. Output only a float score using standard ASCII
characters.
\end{quote}

\noindent\textbf{Similarity-Judge User Turn} (template):
\begin{quote}
\small\ttfamily
Compare the semantic similarity of the following two texts. Score them from
0.0 to 1.0, where 1.0 means they express the exact same ideas and
conclusions, and 0.0 means they are completely unrelated or contradictory.
Return ONLY the float value.\newline
\newline
Text~1:\newline
\{response\_1\}\newline
\newline
Text~2:\newline
\{response\_2\}
\end{quote}

\noindent\textbf{Change-Summariser System Prompt:}
\begin{quote}
\small\ttfamily
You are a concise technical diff reviewer. Use only standard ASCII
characters.
\end{quote}

\noindent\textbf{Change-Summariser User Turn} (template):
\begin{quote}
\small\ttfamily
You are comparing two versions of an AI-generated project plan. Describe in
2--3 bullet points what specifically changed or improved from Version A to
Version B. Focus on structural differences, added/removed tasks, or clarity
improvements.\newline
\newline
Version A:\newline
\{prev\}\newline
\newline
Version B:\newline
\{curr\}
\end{quote}

\subsection{Method~1 v2 — Self-Critique and Refinement (M1\,v2)}
\label{sec:appendix-m1v2}

\textbf{Task:} Scenario~A.  Each iteration consists of three LLM calls:
initial generation (\texttt{temperature=0.7}), critique, and refinement.

\medskip
\noindent\textbf{Critique System Prompt:}
\begin{quote}
\small\ttfamily
You are a pedantic quality auditor. Identify 3 specific flaws.
\end{quote}

\noindent\textbf{Critique User Turn} (template):
\begin{quote}
\small\ttfamily
Original Request: \{prompt\}\newline
\newline
Generated Response:\newline
\{response\}\newline
\newline
---\newline
Critique the response above. Identify exactly 3 potential inaccuracies, vague
statements, or missed requirements from the original request. Be specific and
critical. Use only standard ASCII characters.
\end{quote}

\noindent\textbf{Refinement System Prompt:}
\begin{quote}
\small\ttfamily
You are a perfectionist editor. Correct all flaws and deliver a premium
response.
\end{quote}

\noindent\textbf{Refinement User Turn} (template):
\begin{quote}
\small\ttfamily
Original Request: \{prompt\}\newline
\newline
Initial Draft:\newline
\{response\}\newline
\newline
Critique of Draft:\newline
\{critique\}\newline
\newline
---\newline
Generate a final, polished version of the response that corrects all the
flaws identified in the critique and ensures 100\% alignment with the
original request. Use only standard ASCII characters.
\end{quote}

\subsection{Method~2 — Decomposed Prompting (M2)}
\label{sec:appendix-m2}

\textbf{Task:} Scenario~A.  Two sequential LLM calls: \emph{fact extraction}
then \emph{synthesis}.

\medskip
\noindent\textbf{Fact-Extraction System Prompt:}
\begin{quote}
\small\ttfamily
You are an expert fact extractor. Given a task, list out the concrete
factual requirements or data points needed to solve it. Respond with bullet
points only.
\end{quote}

\noindent\textbf{Synthesis System Prompt:}
\begin{quote}
\small\ttfamily
You are an expert synthesizer. Your goal is to generate a comprehensive
response to the Original Request. Use the `Facts extracted' as the primary
technical source of truth, but ensure you also address EVERY specific
requirement mentioned in the Original Request (e.g., task structure, story
points, teams, diagrams). Use only standard ASCII characters. Do not
hallucinate information not supported by the facts or the request.
\end{quote}

\noindent\textbf{Synthesis User Turn} (template):
\begin{quote}
\small\ttfamily
Original Request: \{original\_prompt\}\newline
\newline
Facts extracted from current reasoning:\newline
\{facts\}\newline
\newline
Please generate the final, high-quality response that meets all criteria.
\end{quote}

\subsection{Method~2 v2 — Context-Aware Synthesis (M2\,v2)}
\label{sec:appendix-m2v2}

\textbf{Task:} Scenario~A.  Identical to M2 except the original prompt is
injected into the synthesis step as an explicit requirements checklist,
preventing information loss between the two stages.

\medskip
\noindent\textbf{Enhanced Synthesis User Turn} (template):
\begin{quote}
\small\ttfamily
Original Request (USE AS CHECKLIST --- ensure every requirement below is
addressed):\newline
\{original\_prompt\}\newline
\newline
Facts extracted from current reasoning:\newline
\{facts\}\newline
\newline
Please generate the final, high-quality response that meets ALL criteria
listed in the original request. Use only standard ASCII characters.
\end{quote}

\subsection{Method~3 — Sequential Task Agents (M3)}
\label{sec:appendix-m3}

\textbf{Task:} Scenario~B.  Four single-task agents run sequentially; each
receives the original scenario plus the outputs of prior agents.

\medskip
\noindent\textbf{Baseline Multi-Task System Prompt:}
\begin{quote}
\small\ttfamily
You are a senior Site Reliability Engineer (SRE) performing incident
response. You have four responsibilities:\newline
1.~Identify the probable root cause from the symptoms provided.\newline
2.~Rank all affected components by severity (Critical / High / Medium / Low).\newline
3.~Provide a step-by-step remediation plan including rollback steps if any
action worsens the situation.\newline
4.~Draft a concise post-mortem report with timeline, impact, root cause, and
action items.\newline
Return all four sections in a single structured document.
\end{quote}

\noindent\textbf{Agent~1 — Root-Cause Analysis System Prompt:}
\begin{quote}
\small\ttfamily
You are a specialized root cause analysis agent. Your ONLY task is to
identify the single most probable root cause of the incident described. State
it in 2--4 sentences. Do not provide remediation or post-mortem content.
\end{quote}

\noindent\textbf{Agent~2 — Severity Ranking System Prompt:}
\begin{quote}
\small\ttfamily
You are a specialized incident severity ranking agent. Your ONLY task is to
rank the affected components listed in the scenario by severity: Critical,
High, Medium, or Low. Format your output as a markdown table with columns:
Component | Severity | Reason. Do not provide remediation steps.
\end{quote}

\noindent\textbf{Agent~3 — Remediation Plan System Prompt:}
\begin{quote}
\small\ttfamily
You are a specialized remediation planning agent. Your ONLY task is to
provide a numbered step-by-step remediation plan for the incident. For any
action that might worsen the situation, include a rollback step immediately
after. Do not re-diagnose the incident.
\end{quote}

\noindent\textbf{Agent~4 — Post-Mortem System Prompt:}
\begin{quote}
\small\ttfamily
You are a specialized post-mortem writing agent. Your ONLY task is to draft a
concise post-mortem report. It must contain these sections: Timeline, Impact
Summary, Root Cause, Remediation Applied, and Action Items to prevent
recurrence. Keep each section brief and factual.
\end{quote}

\subsection{Method~3 v2 — Consensus Reconciliation (M3\,v2)}
\label{sec:appendix-m3v2}

\textbf{Task:} Scenario~B.  Adds a fifth \emph{Reconciler} agent after the
four agents above.

\medskip
\noindent\textbf{Reconciler System Prompt:}
\begin{quote}
\small\ttfamily
You are a high-level SRE director ensuring cross-team consistency.
\end{quote}

\noindent\textbf{Reconciler User Turn} (template):
\begin{quote}
\small\ttfamily
Scenario: \{prompt\}\newline
\newline
--- AGENT OUTPUTS ---\newline
ROOT CAUSE: \{root\_cause\}\newline
SEVERITY: \{severity\}\newline
REMEDIATION: \{remediation\}\newline
POST-MORTEM: \{post\_mortem\}\newline
\newline
---\newline
You are the Lead SRE Reconciler. Your task is to review the independent agent
outputs above for any technical contradictions or inconsistencies (e.g., if
remediation does not match the root cause).\newline
\newline
Perform these steps:\newline
1.~Identify any discrepancies.\newline
2.~Resolve them by creating a single, authoritative, and consistent incident
report.\newline
3.~Ensure the final report follows the 4-section structure (Root Cause,
Severity, Remediation, Post-Mortem).\newline
Use only standard ASCII characters.
\end{quote}

\subsection{Method~4 — Enhanced Data Registry (M4)}
\label{sec:appendix-m4}

\textbf{Task:} Scenario~C (short diagnostic query).  The same diagnostic
question is posed twice — once with sparse raw data and once with an enriched
Virtual Registry.

\medskip
\noindent\textbf{Diagnostic Query (both runs):}
\begin{quote}
\small\ttfamily
Analyze the data provided. Why might the system be blowing warm air, and how
does the current state of CMP-1 relate to VLV-3? Provide a short diagnosis.
\end{quote}

\noindent\textbf{Diagnostic System Prompt (both runs):}
\begin{quote}
\small\ttfamily
You are a diagnostic AI. Analyze the provided system data and the user query.
Output a 3-sentence diagnosis.
\end{quote}

\noindent\textbf{Run~1 — Standard Data Payload:}
\begin{quote}
\small\ttfamily
[\{"id": "CMP-1", "type": "Compressor", "status": "active", "temp": "high"\},\newline
\ \{"id": "VLV-3", "type": "ExpansionValve", "status": "active", "jammed": "true"\},\newline
\ \{"id": "SNS-9", "type": "TempSensor", "status": "offline", "reading": "null"\}]
\end{quote}

\noindent\textbf{Run~2 — Enhanced Virtual Registry Payload (abbreviated):}
\begin{quote}
\small\ttfamily
\{"System\_Component\_Registry": [\newline
\ \ \{"id": "CMP-1", "type": "Scroll Compressor (5-Ton)",\newline
\ \ \ "operational\_status": "Active / Overheating",\newline
\ \ \ "telemetry": \{"casing\_temp\_F": 210, "nominal\_max\_F": 190\},\newline
\ \ \ "description": "Primary refrigerant compressor. High temperature indicates
potential liquid slugging or loss of charge."\},\newline
\ \ \{"id": "VLV-3", "type": "Thermostatic Expansion Valve (TXV)",\newline
\ \ \ "operational\_status": "Active / Faulty",\newline
\ \ \ "telemetry": \{"position": "stuck\_closed", "superheat\_F": 45\},\newline
\ \ \ "description": "Regulates refrigerant flow. Jammed closed state causes
high superheat and compressor overheating."\},\newline
\ \ \{"id": "SNS-9", "type": "Evaporator Discharge Air Temp Sensor",\newline
\ \ \ "operational\_status": "Offline",\newline
\ \ \ "telemetry": \{"reading": null, "last\_known\_reading": 55\},\newline
\ \ \ "description": "Measures supply air temp. Offline status prevents the
system controller from modulating airflow correctly."\}]\}
\end{quote}

\subsection{Method~5 — Static Glossary Injection (M5)}
\label{sec:appendix-m5}

\textbf{Task:} Scenario~C (full BMS fault description).  Two runs: without
and with a pre-defined domain glossary prepended to the system prompt.

\medskip
\noindent\textbf{Run~1 — No-Glossary System Prompt:}
\begin{quote}
\small\ttfamily
You are an AI assistant helping a building systems engineer. Answer the query
directly and concisely.
\end{quote}

\noindent\textbf{Run~2 — With-Glossary System Prompt (structure):}
\begin{quote}
\small\ttfamily
You are an AI assistant helping a building systems engineer. Answer the query
directly and concisely. Use the following domain glossary to interpret all
acronyms precisely:\newline
\newline
{[}DOMAIN GLOSSARY: Building Systems \& HVAC{]}\newline
- BMS: Building Management System\newline
- AHU: Air Handling Unit\newline
- MAT: Mixed Air Temperature\newline
- SAT: Supply Air Temperature\newline
- VAV: Variable Air Volume (terminal unit)\newline
- DDC: Direct Digital Controller\newline
- VFD: Variable Frequency Drive\newline
- CFM: Cubic Feet per Minute (airflow volume)\newline
- RTU: Rooftop Unit\newline
- CO\textsubscript{2}: Carbon Dioxide (indoor air quality indicator)\newline
- OSA: Outside Air\newline
- BAS: Building Automation System\newline
- DX: Direct Expansion (refrigerant-based cooling)\newline
- OAT: Outside Air Temperature\newline
- SP: Setpoint\newline
- EWT: Entering Water Temperature\newline
- LWT: Leaving Water Temperature\newline
- FCU: Fan Coil Unit\newline
- CV: Constant Volume\newline
- RA: Return Air
\end{quote}

\subsection{Method~5 v2 — Dynamic Glossary Injection (M5\,v2)}
\label{sec:appendix-m5v2}

\textbf{Task:} Scenario~C.  Adds an LLM-based term-selection step before
injection so that only contextually relevant acronyms are included.

\medskip
\noindent\textbf{Term-Identification System Prompt:}
\begin{quote}
\small\ttfamily
You are a precise technical indexer. Output only comma-separated terms.
\end{quote}

\noindent\textbf{Term-Identification User Turn} (template):
\begin{quote}
\small\ttfamily
Prompt: \{prompt\}\newline
\newline
Glossary Terms Available: \{terms\_list\}\newline
\newline
Identify which of the acronyms above are used in the prompt. Return ONLY a
comma-separated list of the terms.
\end{quote}

\noindent\textbf{Final System Prompt (template, relevant terms only):}
\begin{quote}
\small\ttfamily
You are an AI assistant helping a building systems engineer. Answer the query
directly and concisely. Use the following RELEVANT domain glossary entries to
interpret acronyms precisely:\newline
\newline
\{injected\_glossary\}
\end{quote}

\subsection{LLM-as-Judge Evaluation Prompt}
\label{sec:appendix-eval}

All method outputs were evaluated against their respective baselines using the
following rubric.  Results are returned as a plain-text \texttt{SCORE} line
followed by a one-sentence \texttt{REASON}.

\medskip
\noindent\textbf{Evaluator System Prompt:}
\begin{quote}
\small\ttfamily
You are an expert AI evaluator. Use only standard ASCII characters in your
response (no arrows, special dashes, or non-breaking spaces). Compare two
AI-generated responses (Control vs Method) based on the following metrics:\newline
1.~Accuracy / Hallucinations (Does it invent false facts or stick to the
truth?)\newline
2.~Clarity \& Structure (Is it easier to read and follow?)\newline
3.~Directness (Does it directly answer the objective without fluff?)\newline
\newline
Determine an overall score. If the Method response is superior, return
exactly the word ``Better''. If the Control response is superior, return
exactly the word ``Worse''. If they are roughly equal in quality or
identically flawed, return exactly the word ``Same''.\newline
\newline
You MUST return your response in the following format:\newline
SCORE: {[}Better / Worse / Same{]}\newline
REASON: {[}1 concise sentence explaining why you gave this score{]}
\end{quote}

\noindent\textbf{Evaluator User Turn} (template):
\begin{quote}
\small\ttfamily
Objective:\newline
\{objective\}\newline
\newline
--- CONTROL RESPONSE ---\newline
\{control\_response\}\newline
\newline
--- METHOD RESPONSE ---\newline
\{method\_response\}
\end{quote}